\documentclass[]{aa}
\usepackage{graphics}
\usepackage{graphicx}
\usepackage{txfonts}

\def\apj{ApJ}                 
                
\def\apjs{ApJS}

\def\aap{A\&A}

\def\mnras{MNRAS}

\def\pasj{PASJ}

\def\msol{M$_{\odot}$}
\def\lsol{L$_{\odot}$}
\begin{document}      
%
   \title{The CoRoT primary target HD~52265: models and seismic tests.} 
%
 
        \titlerunning{Seismic modelling of HD~52265}  
 
   \author{
M.~Soriano
\and S.~Vauclair 
\and G.~Vauclair
\and M.~Laymand   
}
 
   \offprints{M. Soriano}

   \institute{Laboratoire d'Astrophysique de Toulouse et Tarbes - UMR 5572 - Universit\'e Paul Sabatier Toulouse III - CNRS, 14, av. E. Belin, 31400 Toulouse, France} 

\mail{sylvie.vauclair@ast.obs-mip.fr}
   
\date{Received \rule{2.0cm}{0.01cm} ; accepted \rule{2.0cm}{0.01cm} }

\authorrunning{ Soriano et al.}

\abstract
 {}
{ HD~52265 is the only known exoplanet-host star selected as a main target for the seismology programme of the CoRoT satellite. As such, it will be observed continuously during five months, which is of particular interest in the framework of planetary systems studies. This star was misclassified as a giant in the Bright Star Catalog, while it is more probably on the main-sequence or at the beginning of the subgiant branch. We performed an extensive analysis of this star, showing how asteroseismology may lead to a precise determination of its external parameters and internal structure.}
{We first reviewed the observational constraints on the metallicity, the gravity and the effective temperature derived from the spectroscopic observations of HD~52265. We also derived its luminosity using the Hipparcos parallax. We computed the evolutionary tracks for models of various metallicities which cross the relevant observational error boxes in the gravity-effective temperature plane. We selected eight different stellar models which satisfy the observational constraints, computed their p-modes frequencies and analysed specific seismic tests.}
{The possible models for HD~52265, which satisfy the constraints derived from the spectroscopic observations, are different in both their external and internal parameters. They lie either on the main sequence or at the beginning of the subgiant branch. The differences in the models lead to quite different properties of their oscillation frequencies. We give evidences of an interesting specific behaviour of these frequencies in case of helium-rich cores: the ``small separations'' may become negative and give constraints on the size of the core. We expect that the observations of this star by the CoRoT satellite will allow choosing between these possible models. }
{}


\maketitle
                                                                                                                                         
\section{Introduction} 

The study of the internal structure of exoplanet-host stars (EHS) is a key issue  for the understanding of the process of planetary systems formation. In this context, asteroseismology technics represents an excellent tool to help determining the structural differences between stars with and without detected planets. The most stricking among these differences is the observed overmetallicity of EHS compared to stars without planets (Santos et al. \cite{santos03} and \cite{santos05}, Gonzalez \cite{gonzalez03}, Fischer \& Valenti \cite{fischer05}). This overmetallicity may be of primordial origin, which means that it was already present in the interstellar cloud out of which the stellar system emerged, or related to accretion of hydrogen poor material onto the star during the early phases of the planetary system formation (Bazot \& Vauclair \cite{bazot04}). Although the first scenario seems more realistic and probable, the second one is not yet completely excluded. Note that the stars for which no planet have been detected may very well possess planets with large orbits, like the Sun. The so-called exoplanets-host stars correspond to those for which the giant planets have migrated towards the central star during the early phases. One must keep in mind that the differences between the two cases (stars observed with or without planets) are related to the phenomenon of planet migration more than that of planet formation, even if both effects are connected.

Besides the question of the origin of the overmetallicity in EHS, recent papers (Bazot et al. \cite{bazot05}, Laymand \& Vauclair \cite{laymand07}) showed how asteroseismology could lead to more precise determinations of the outer stellar parameters (log $g$, log $T_{eff}$, metallicity) than spectroscopy. Meanwhile, Laymand \& Vauclair \cite{laymand07} showed the importance of computing very precise models of the studied star, taking into account precise metallicity, to infer mass and age. Determination of these parameters by extrapolation in published grids can lead to wrong values.

The EHS $\mu$ Arae was observed with the HARPS spectrometer in June 2004 to obtain radial velocity time series. Their analysis led to the discovery of up to 43 frequencies which could be identified with p-modes of degree $\ell$=0 to 3 (Bouchy et al. \cite{bouchy05}) and a detailed modelling was given in Bazot et al. \cite{bazot05}. Two other very promising EHS for such a seismic analysis are the solar type stars $\iota$ Hor (HD 17051, HR 810) and HD~52265. The case of $\iota$ Hor, which has been observed with HARPS in November 2006, was discussed by Laymand \& Vauclair (\cite{laymand07}). 

HD~52265, mistakenly classified as a G0III-IV in the Bright Star Catalog, was reclassified as a solar type G0V main sequence star by Butler et al. \cite{butler00}. A Jupiter-mass planet orbiting at 0.5~AU with a period of 119 days was discovered independently by Butler et al. (\cite{butler00}) and Naef et al. (\cite{naef01}). The star overmetallicity has been established by a number of subsequent analyses like Gonzalez et al. (\cite{gonzalez01}), Santos et al. (\cite{santos04}), Fischer \& Valenti (\cite{fischer05}), Takeda et al. (\cite{takeda05}) and Gillon \& Magain (\cite{gillon06}). HD~52265 will be one of the main targets of the seismology programme of the CoRoT space mission (Baglin \cite{baglin03}), which has two main scientific programmes: the asteroseismological study of bright variable stars of different types in the H-R diagram and the search for planetary transits. The study of HD~52265 offers a unique opportunity to link the two scientific programmes of CoRoT (Vauclair et al. \cite{vauclair06}).

\begin{table}
\caption{Effective temperatures, gravities and metal abundances observed for HD~52265. [Fe/H] ratios are given in dex. The references for the given values are: Gonzalez et al. (2001) (GLTR01), Santos et al. (2004) (SIM04), Fischer \& Valenti (2005) (FV05), Takeda et al. (2005) (TOSKS05) and Gillon \& Magain (2006) (GM06).}
\label{tab1}
\begin{center}
\begin{tabular}{cccc} \hline
\hline
T$_{\mbox{eff}} (K)$ & log g  & [Fe/H] & Reference \cr
 \hline
6162$\pm$22 & 4.29$\pm$0.04 & 0.27$\pm$0.02 & GLTR01\cr
6103$\pm$52 & 4.28$\pm$0.12 & 0.23$\pm$0.05 & SIM04\cr
6076$\pm$44 & 4.26$\pm$0.06 & 0.19$\pm$0.03 & FV05\cr
6069$\pm$15 & 4.12$\pm$0.03 & 0.19$\pm$0.03 & TOSKS05\cr
6179$\pm$18 & 4.36$\pm$0.03 & 0.24$\pm$0.02 & GM06\cr
\hline
\end{tabular}
\end{center}
\end{table}

\begin{figure*}
\begin{center}
\includegraphics[angle=0,totalheight=5.5cm,width=8cm]{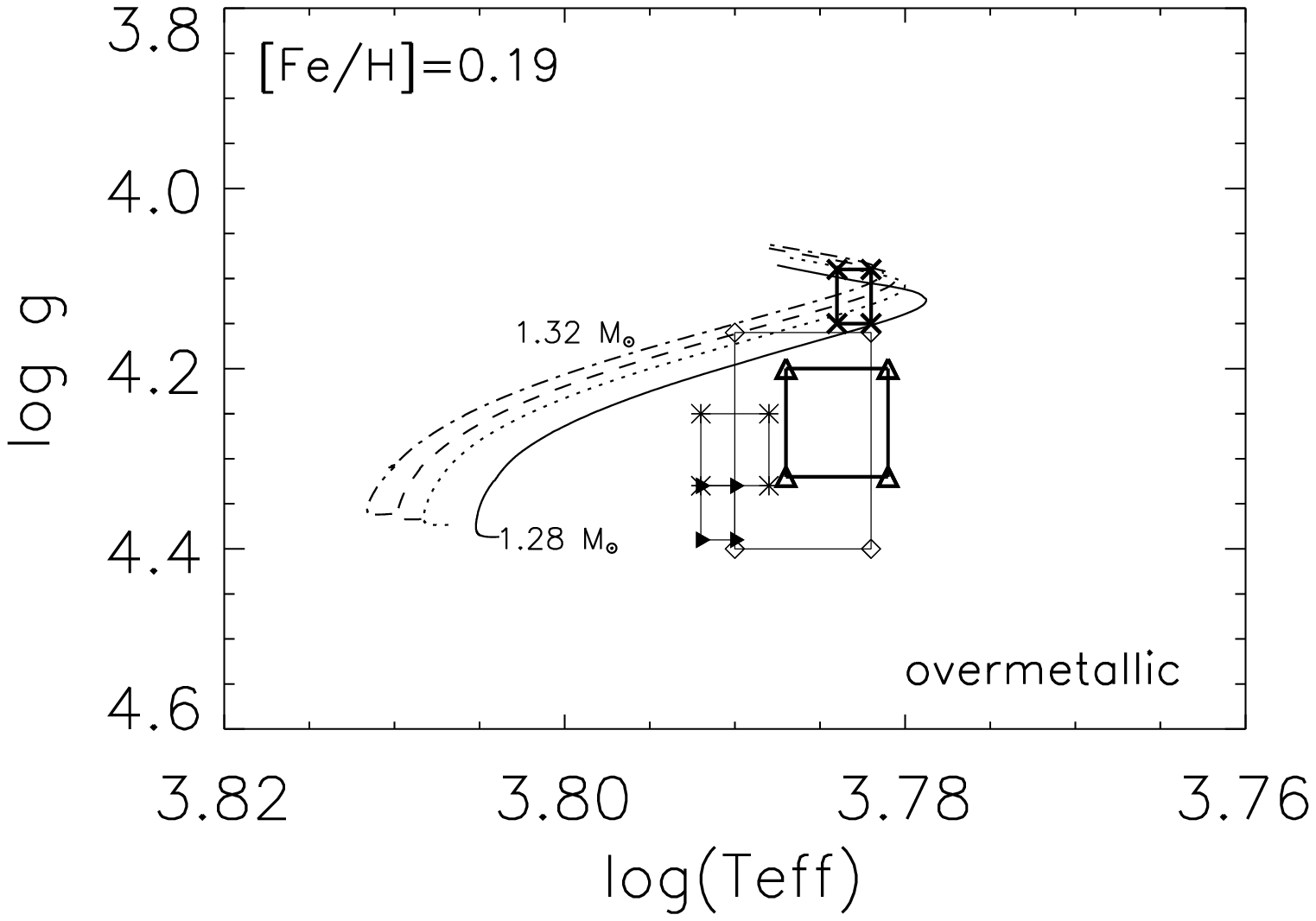}\includegraphics[angle=0,totalheight=5.5cm,width=8cm]{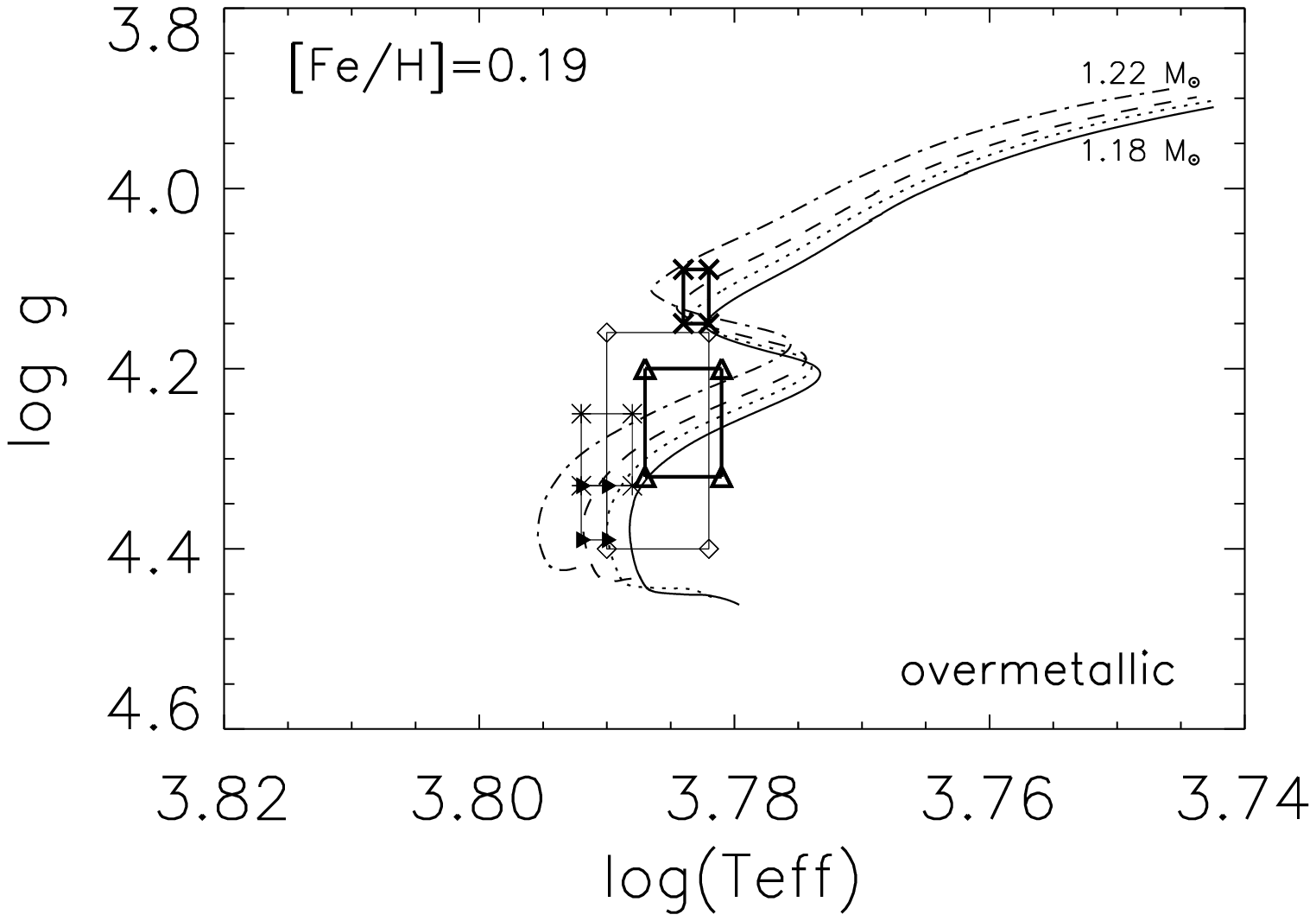}
\includegraphics[angle=0,totalheight=5.5cm,width=8cm]{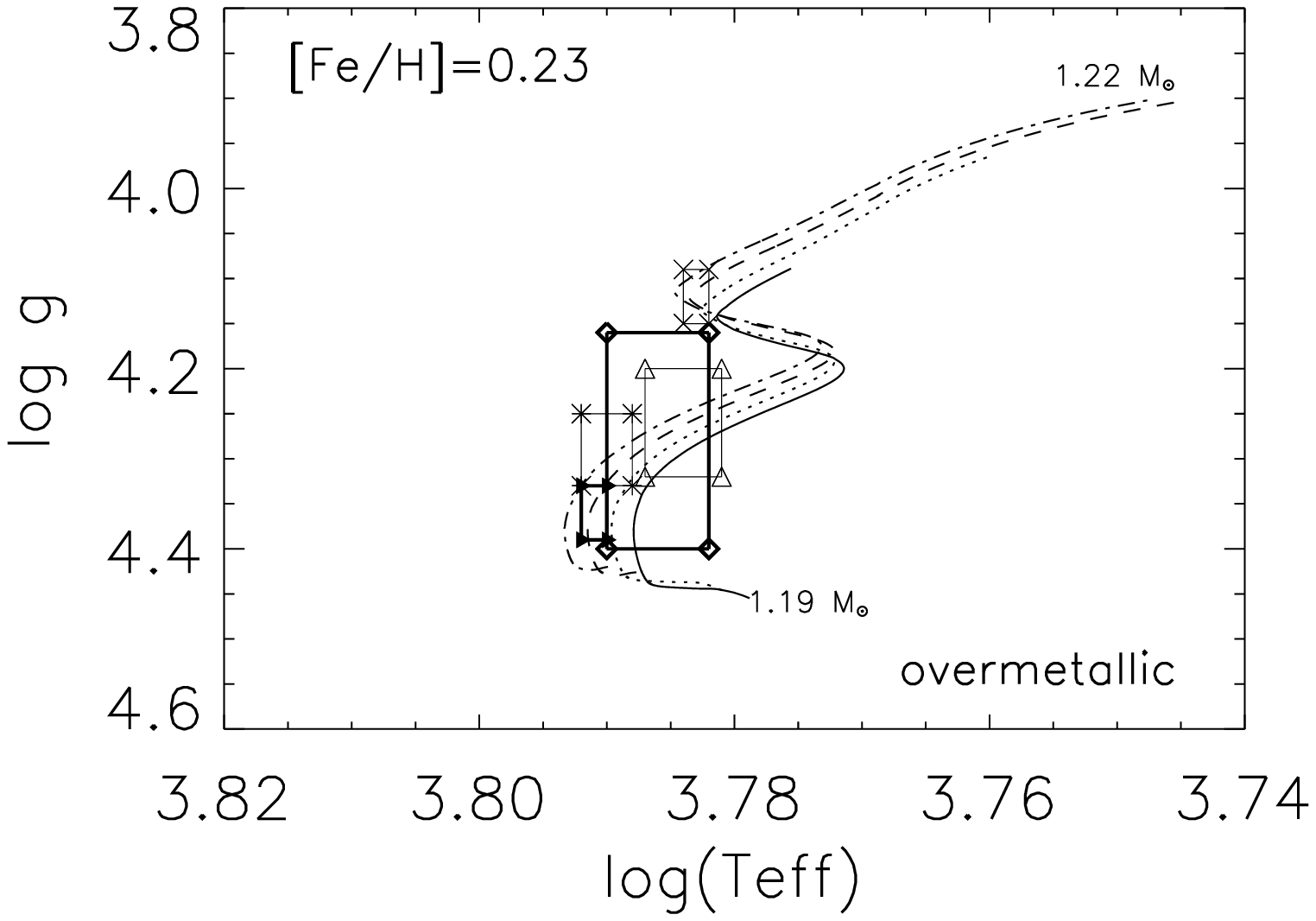}\includegraphics[angle=0,totalheight=5.5cm,width=8cm]{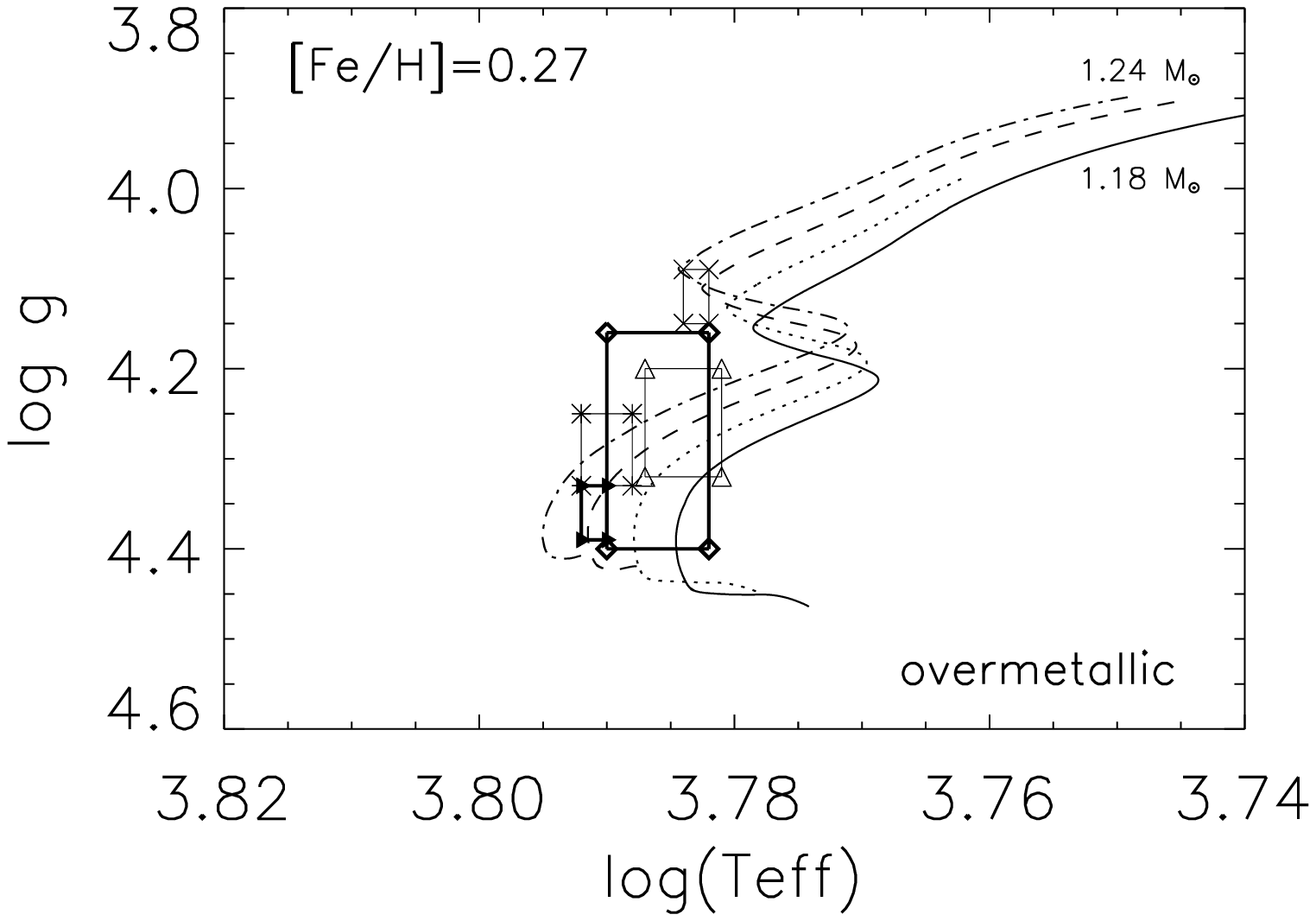}
\includegraphics[angle=0,totalheight=5.5cm,width=8cm]{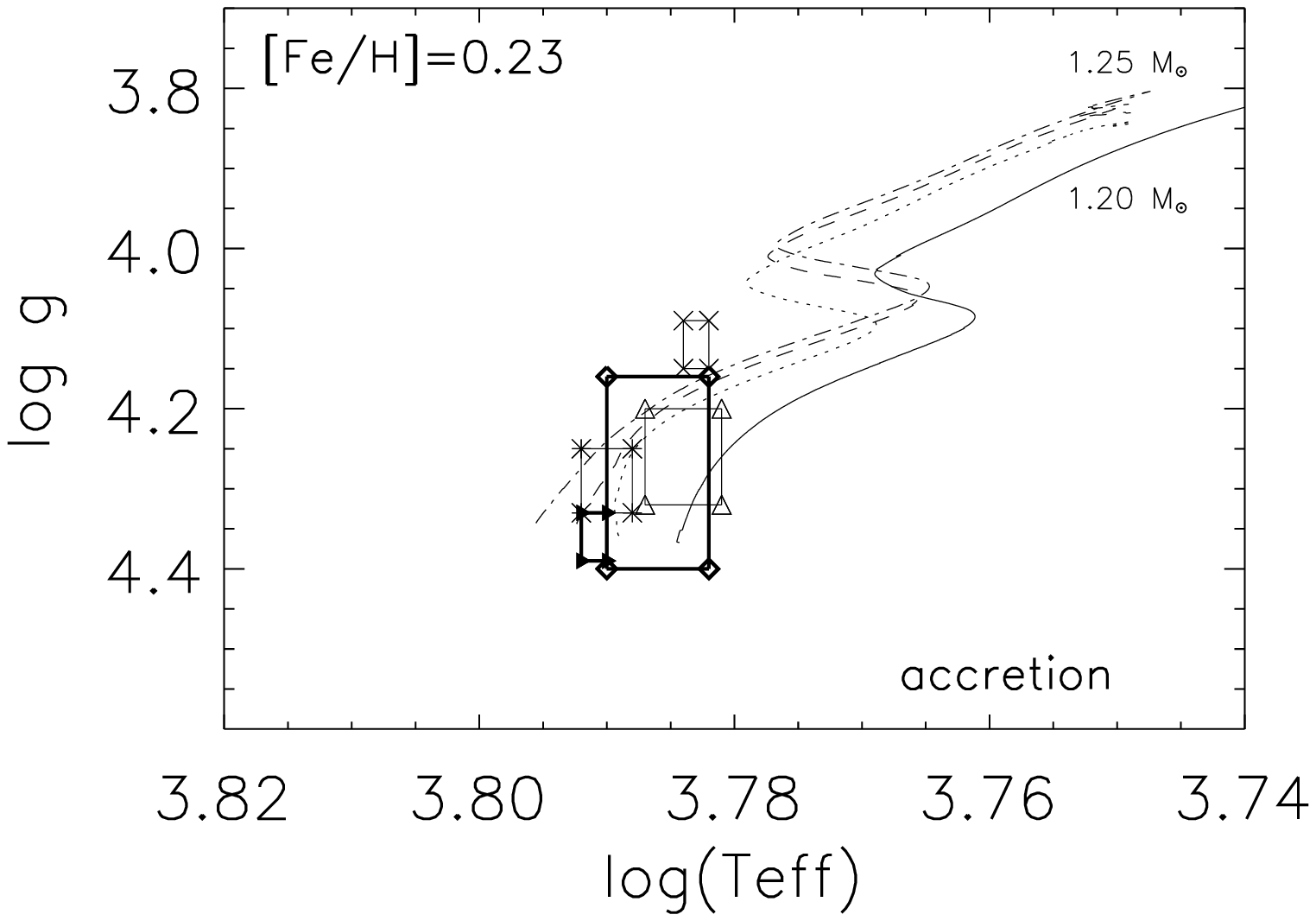}\includegraphics[angle=0,totalheight=5.5cm,width=8cm]{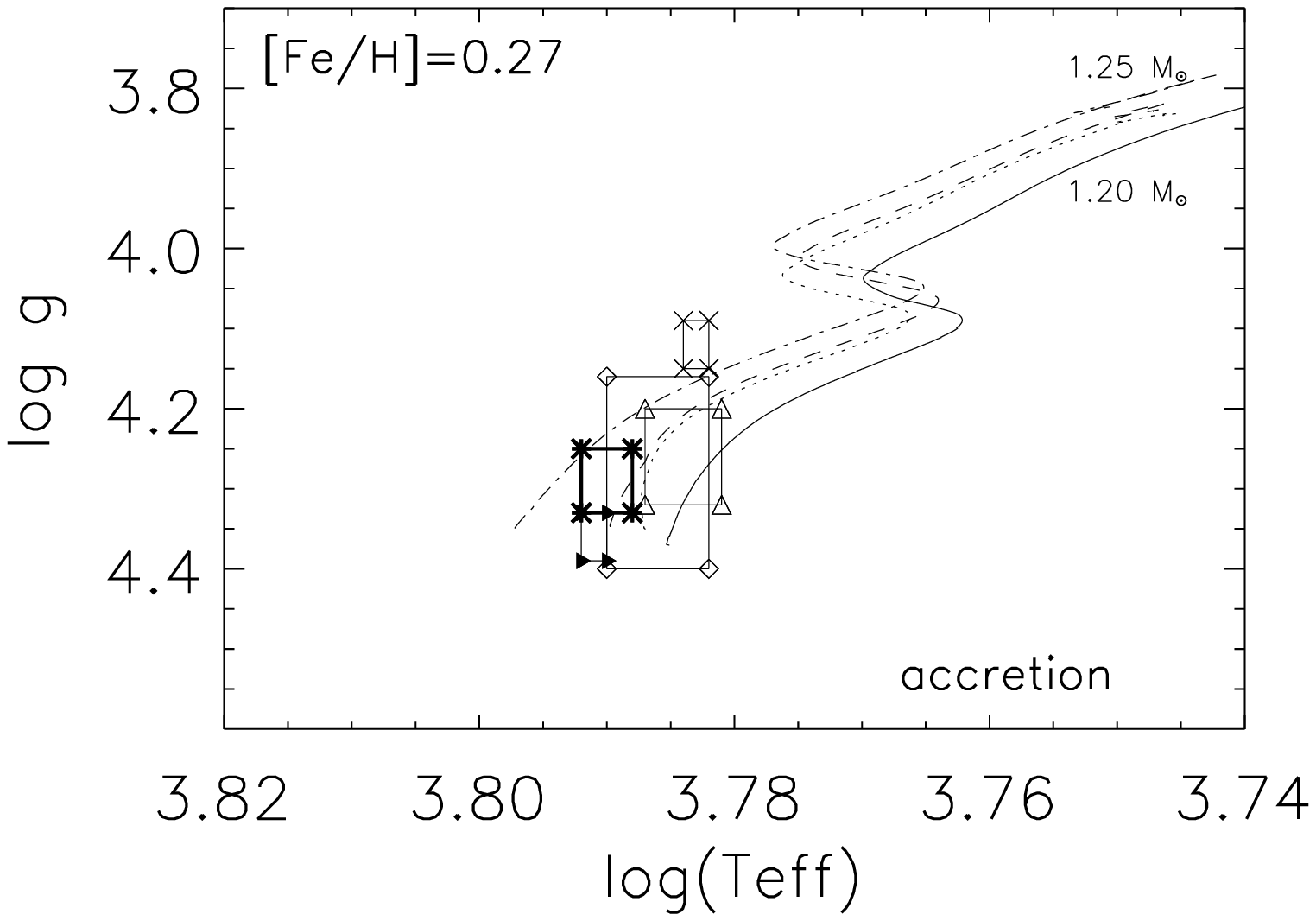}
\end{center}
\caption{This figure shows the constraints on the position of the star HD 52265 in the log~$g$ - log~$T_{eff}$ diagram for the three different metallicities derived from spectroscopic observations. The five error boxes shown are from: Gillon \& Magain (black triangles), Fischer \& Valenti (white triangles), Santos et al. (diamonds), Gonzalez et al. (asterisks), Takeda et al. (crosses). For each selected metallicity, the corresponding authors box is drawn in thick lines. The figures of the upper and middle panels, for [Fe/H]=0.19, 0.23 and 0.27 show the evolutionary tracks for overmetallic models, while the lower figures for [Fe/H]=0.23 and 0.27 are for models with accretion. In the upper-left figure, the evolutionary tracks are for 1.28 \msol (thick line), 1.30 \msol (dotted line), 1.31 \msol (dashed line) and 1.32 \msol (dotted-dashed line) respectively; in the upper-right figure the evolutionary tracks are in the same order for 1.18, 1.19, 1.20 and 1.22 \msol; in the middle-left figure the evolutionary tracks are in the same order for 1.19, 1.20, 1.21 and 1.22 \msol; in the middle-right figure they are in the same order for 1.18, 1.20, 1.22 and 1.24 \msol; in the lower-left figure, the tracks are for 1.20, 1.22, 1.24 and 1.25 \msol; and in the lower-right figure, they are in the same order for 1.20, 1.22, 1.23 and 1.25 \msol.}
\label{fig1}
\end{figure*}

For the present paper we computed evolutionary tracks using [Fe/H] values as given by the various observing groups and selected models lying inside the corresponding error boxes in the log~$g$ - log~$T_{eff}$ plane. We used the TGEC code (Toulouse-Geneva Evolutionary Code) as described in previous papers (e.g. Bazot \& Vauclair \cite{bazot04} and Bazot et al. \cite{bazot05}). We tested the differences obtained in the internal structure of the models for the various possible observed metallicities and the overmetallic versus accretion scenarii. According to the atmospheric parameters derived by the different authors cited above, HD~52265 should have a convective core whose mass largely depends on the adopted metallicity. We computed the p-modes oscillation frequencies for several characteristic models, using the code PULSE (Brassard et al. \cite {brassard92}). We discuss asteroseismic tests which will help choosing among the possible models. We find that, if the star is at the end of the main-sequence or at the beginning of the subgiant branch, the ``small separations'' between modes $\ell=0$ and $\ell=2$ become negative at some frequency, and we specifically discuss this interesting behaviour.

The computations of evolutionary tracks and various modelling of this star are described in Sect.~2. Sect.~3 is devoted to the discussion of asteroseismic tests and predictions for models which lie at the beginning of the main-sequence. In Sect.~4, we specially discuss the case of two models with helium rich cores and we specially emphasize the question of negative small separations. The summary and conclusion are given in Sect.~5

\section{Evolutionary tracks and models}

\subsection{Observational boxes and computations}

Five different groups of observers have determined the metallicity and the external parameters of HD~52265 (see Table~\ref{tab1}). This star has also been observed with the Hipparcos satellite from which the parallax was derived: $\pi=35.63\pm0.84$ mas. The visual magnitude of HD~52265 is given as V = 6.301 (SIMBAD Astronomical data base). From Table~\ref{tab1}, we derive an average value for the effective temperature $T_{eff} = 6115$~K with an uncertainty of order 100~K. Using the tables of Flower (Flower \cite{flower96}), we obtained for the bolometric correction: BC $=-0.03\pm0.01$. With a solar absolute bolometric magnitude of $M_{bol,\odot}=4.75$ (Lejeune et al. \cite{lejeune98}), and taking into account the uncertainty on the parallax, we deduce a luminosity of log~L/L$_{\odot}=0.29\pm0.05$. In any case, we do not use this luminosity as a basic constraint on our models, which are more consistently fitted to the ``triplets'' ([Fe/H], log $g$; log $T_{eff}$) given by spectroscopists.

We computed series of overmetallic and accretion models which could account for the observed parameters of HD~52265. We used the Toulouse-Geneva stellar evolution code, with the OPAL equation of state and opacities (Rogers \& Nayfonov \cite{rogers02}, Iglesias \& Rogers \cite{iglesias96}) and the NACRE nuclear reaction rates (Angulo et al. \cite{angulo99}). In all our models microscopic diffusion was included using the Paquette prescription (Paquette et al. \cite{paquette86}, Michaud et al. \cite{michaud04}). The treatment of convection was done in the framework of the mixing length theory and the mixing length parameter was adjusted as in the Sun ($\alpha = 1.8$). For the overmetallic models, we assumed that helium was enriched as well as metals according to the law given by Isotov \& Thuan (\cite{isotov04}): dY/dZ = $2.8 \pm 0.5$. The accretion models were computed with the same assumptions as in Bazot \& Vauclair (\cite{bazot04}) (instantaneous fall of matter at the beginning of the main sequence and instantaneous mixing inside the convection zone). Neither extra-mixing nor overshoot were taken into account in the present paper.

\begin{figure*}
\begin{center}
\includegraphics[angle=0,totalheight=5.5cm,width=8cm]{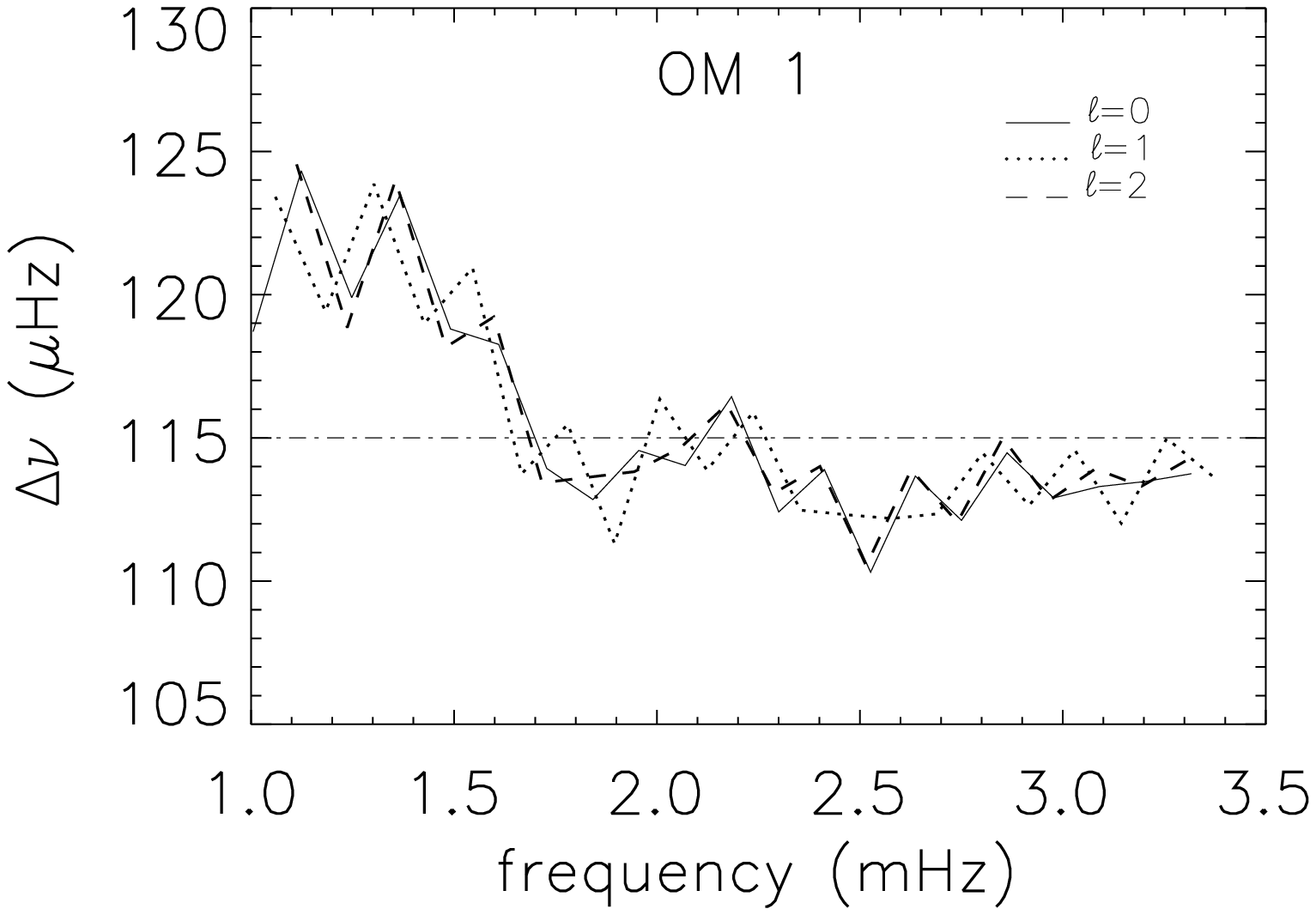}\includegraphics[angle=0,totalheight=5.5cm,width=8cm]{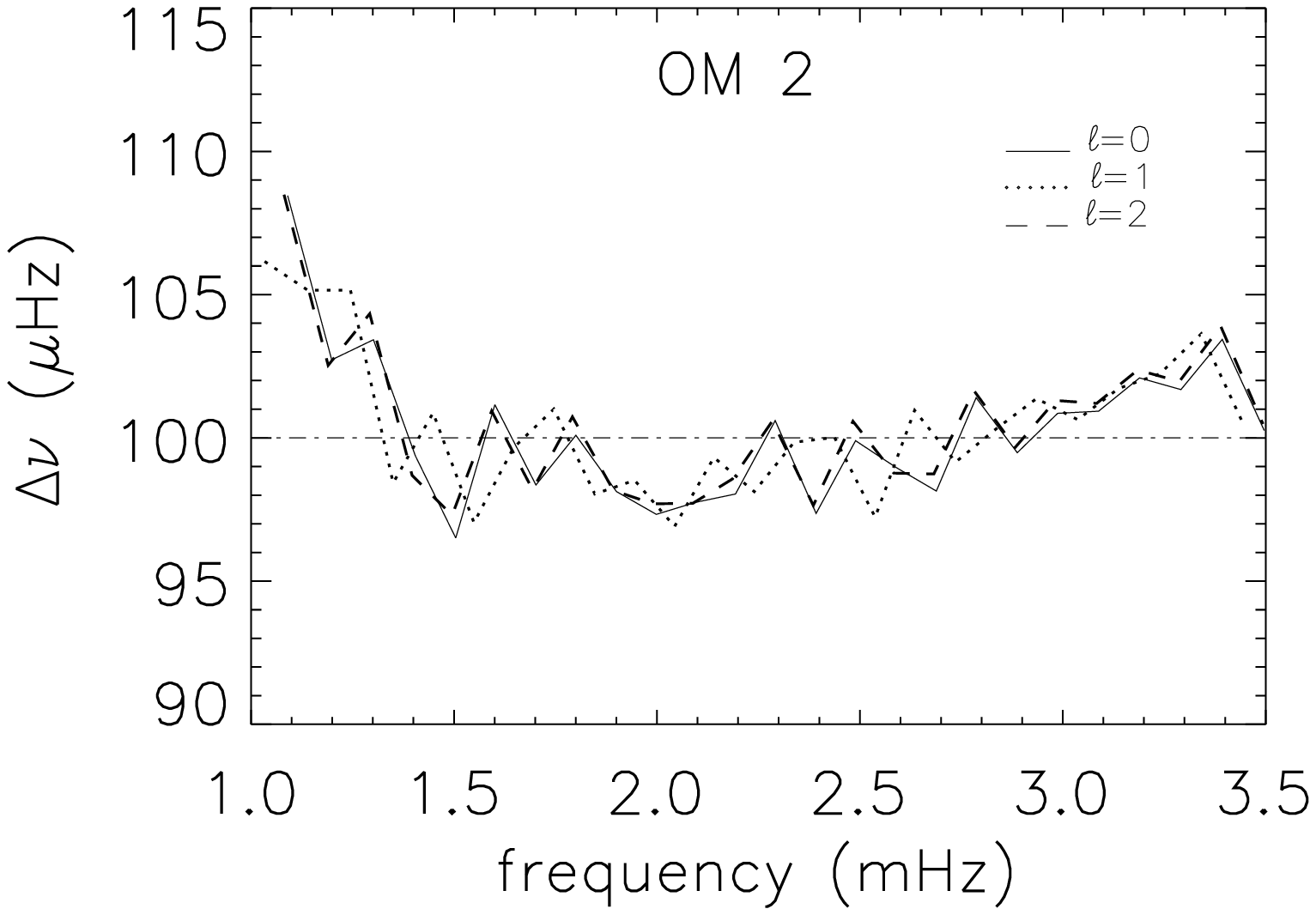}
\includegraphics[angle=0,totalheight=5.5cm,width=8cm]{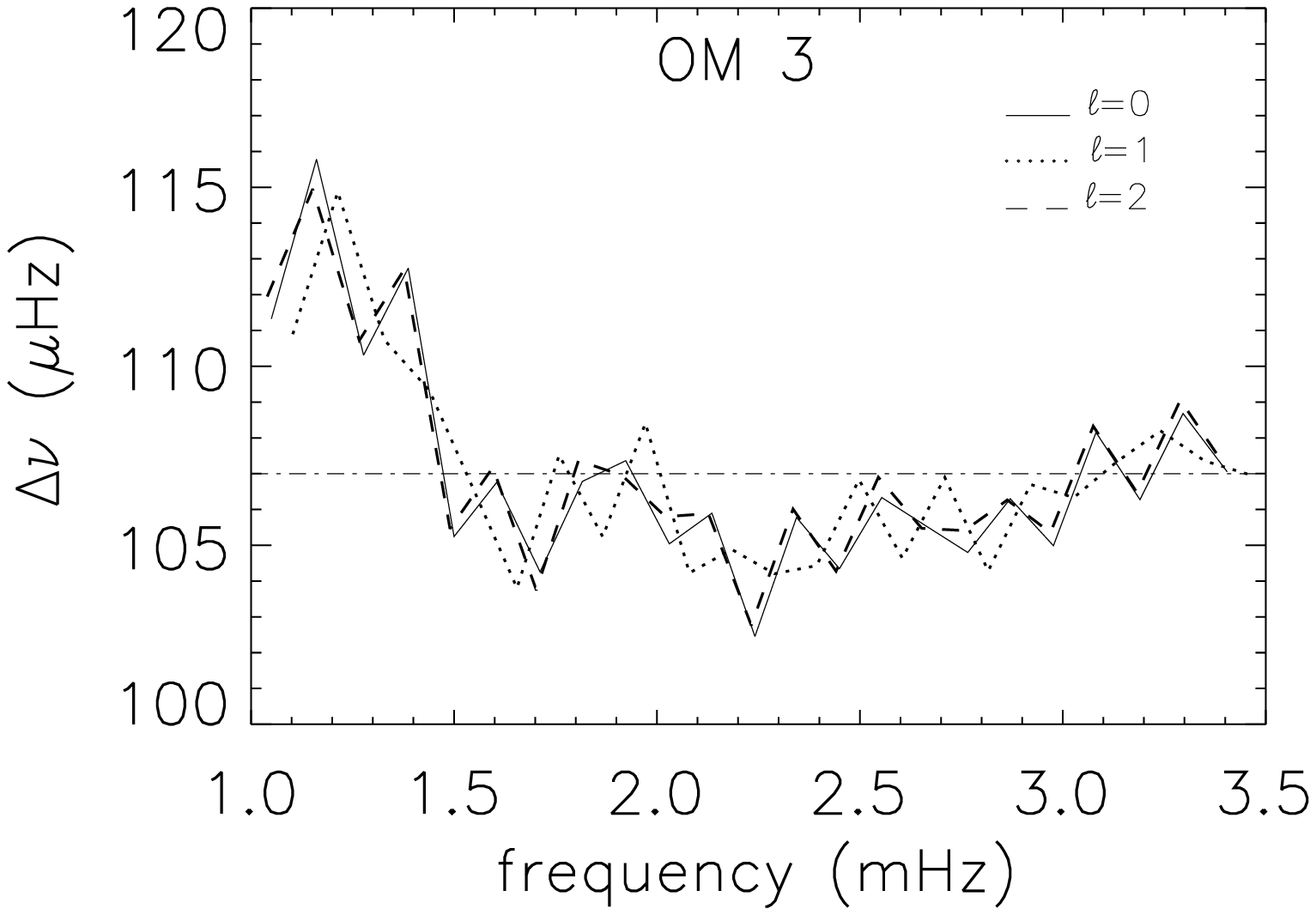}\includegraphics[angle=0,totalheight=5.5cm,width=8cm]{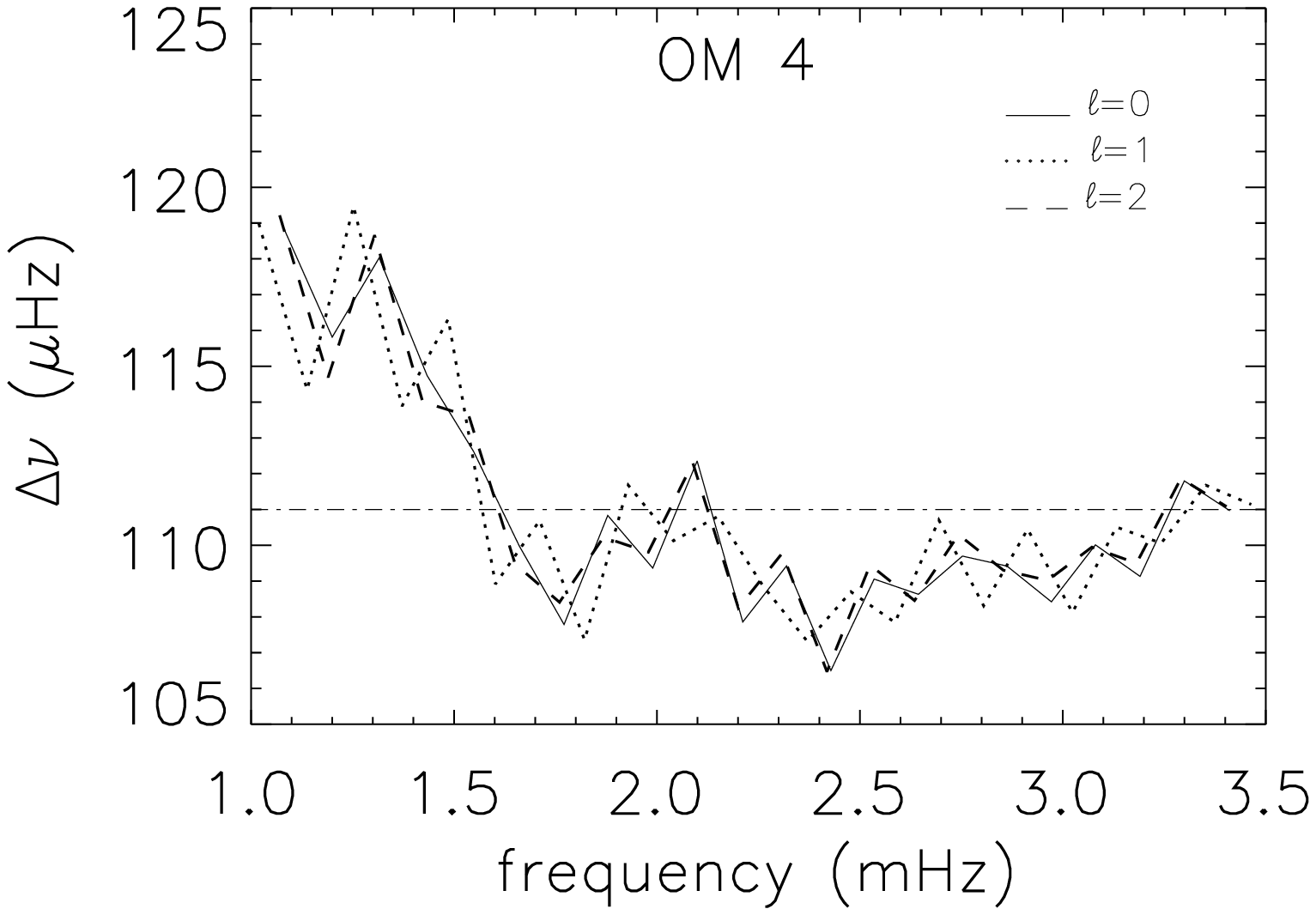}
\includegraphics[angle=0,totalheight=5.5cm,width=8cm]{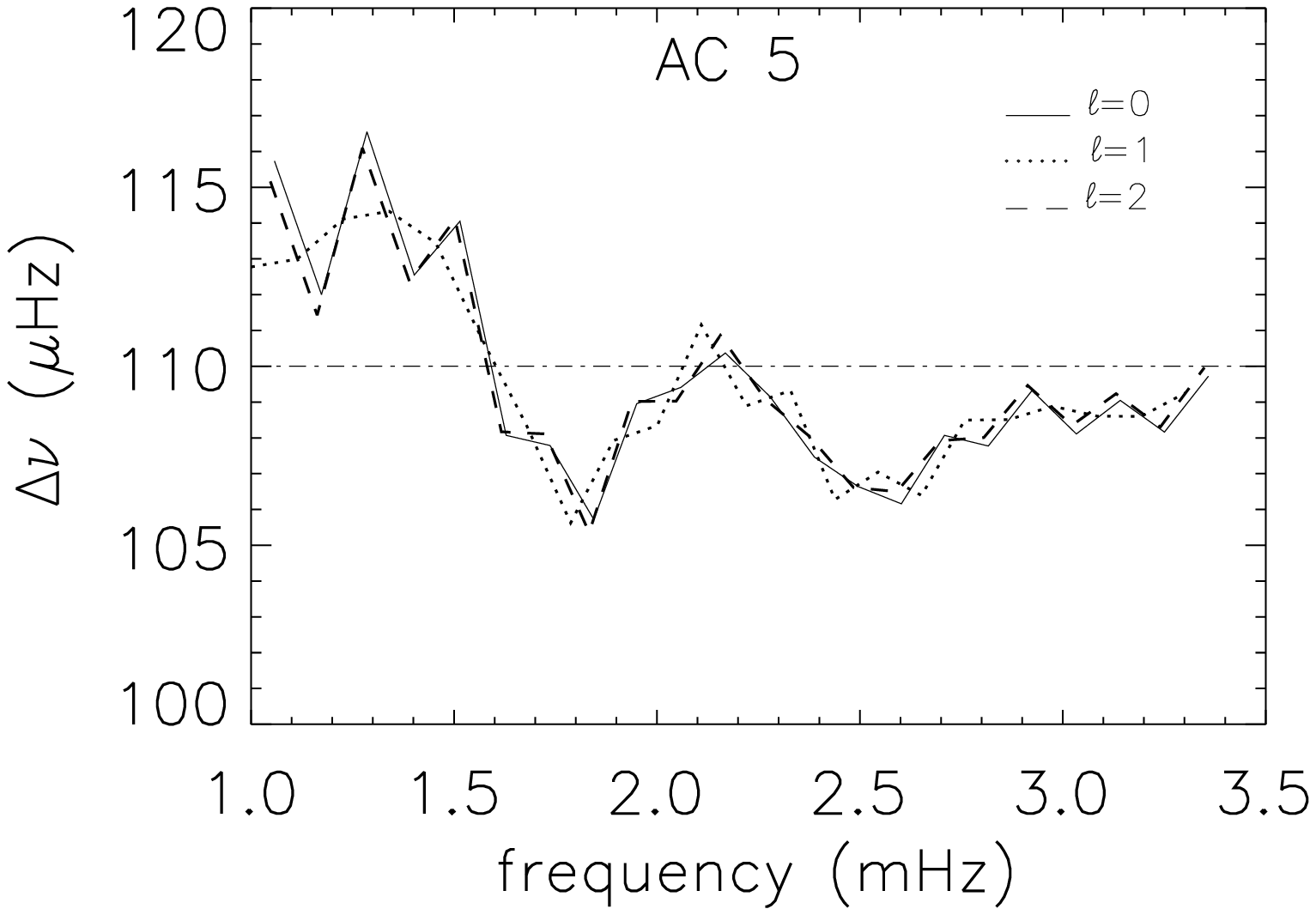}\includegraphics[angle=0,totalheight=5.5cm,width=8cm]{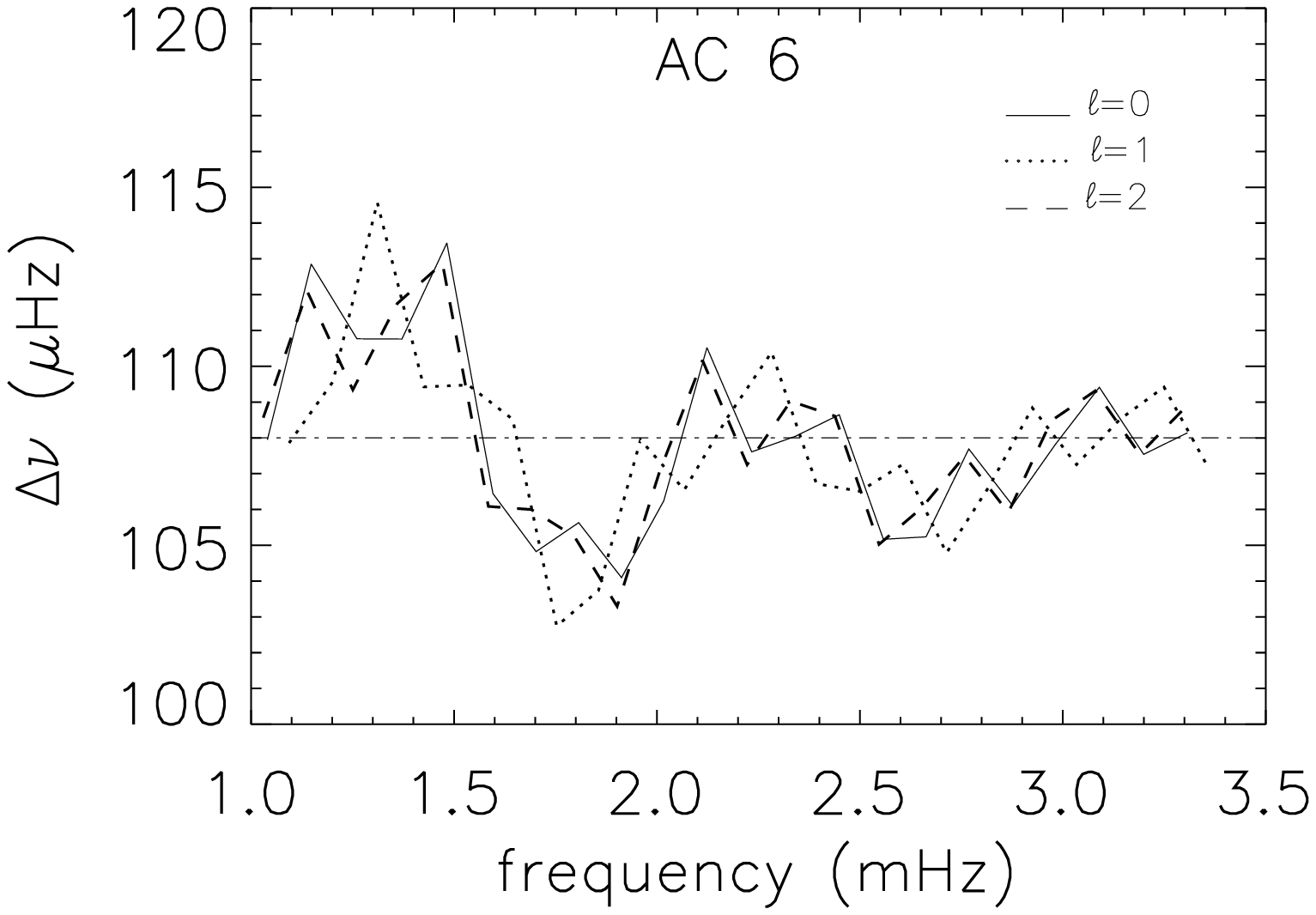}
\end{center}
\caption{Large separations for the overmetallic models 1 to 4 and the accretion models 5 and 6. The horizontal dotted-dashed line correspond to the value of the mean large separation.}
\label{fig2}
\end{figure*}

\begin{figure*}
\begin{center}
\includegraphics[angle=0,totalheight=5.5cm,width=8cm]{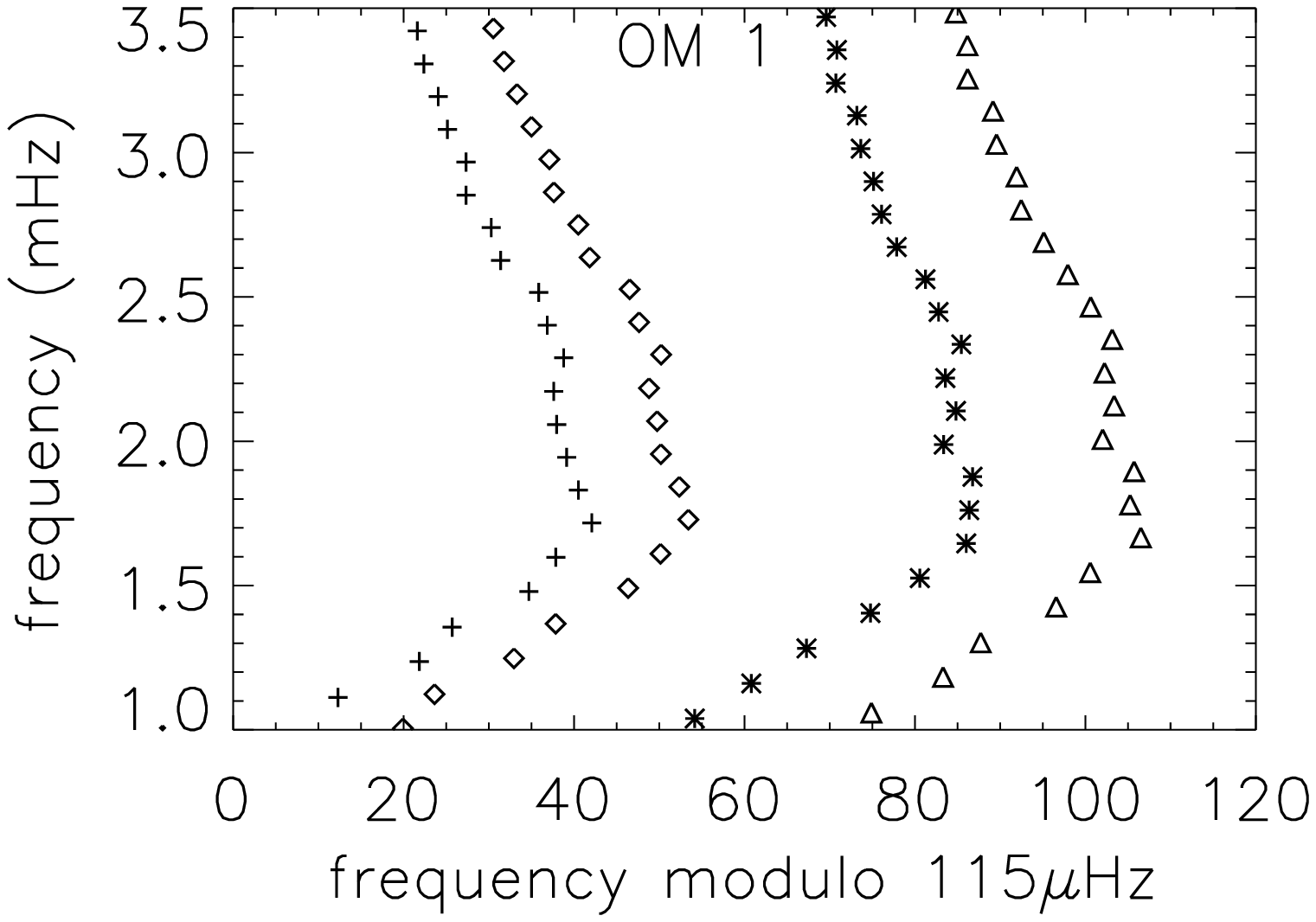}\includegraphics[angle=0,totalheight=5.5cm,width=8cm]{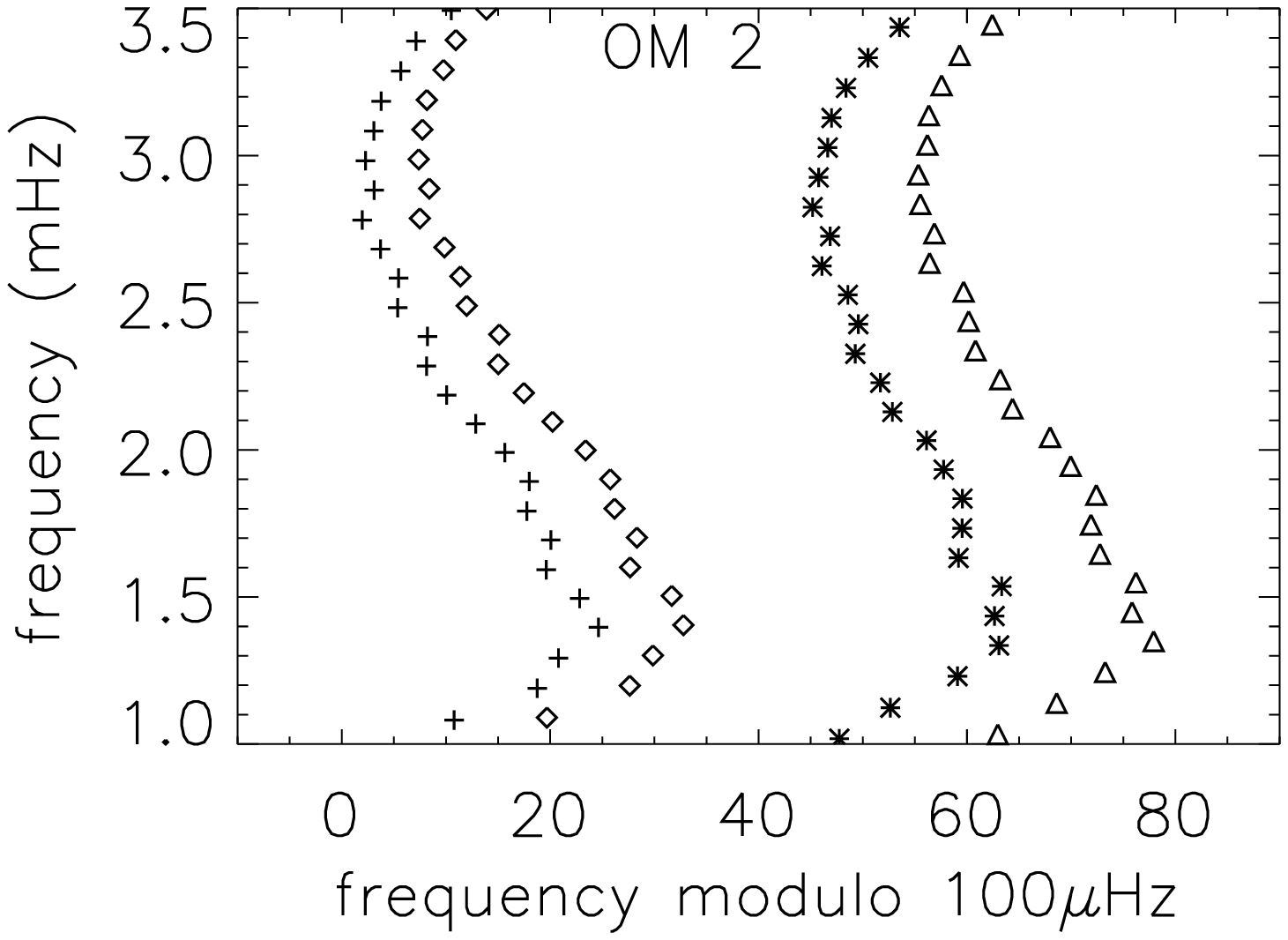}
\includegraphics[angle=0,totalheight=5.5cm,width=8cm]{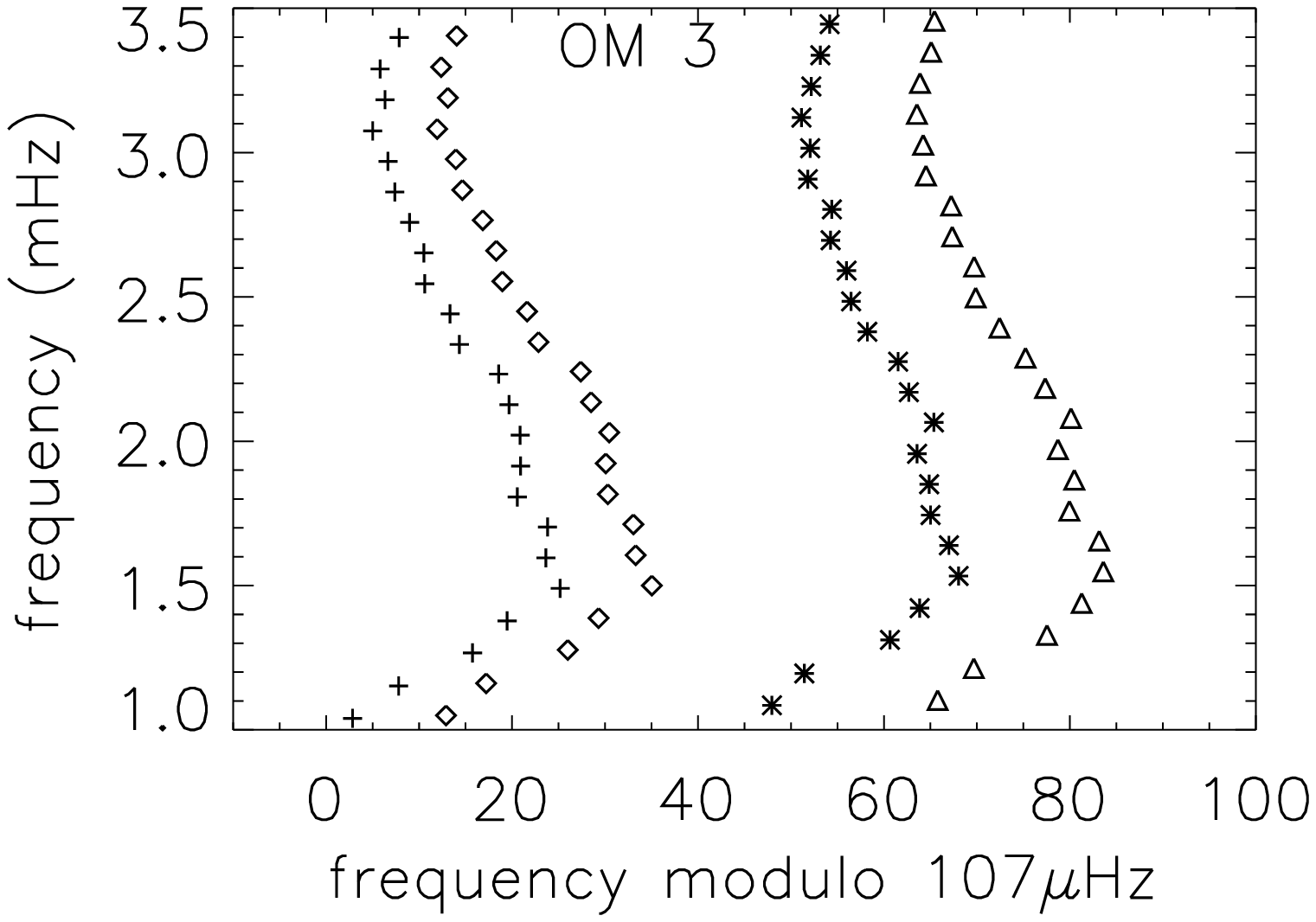}\includegraphics[angle=0,totalheight=5.5cm,width=8cm]{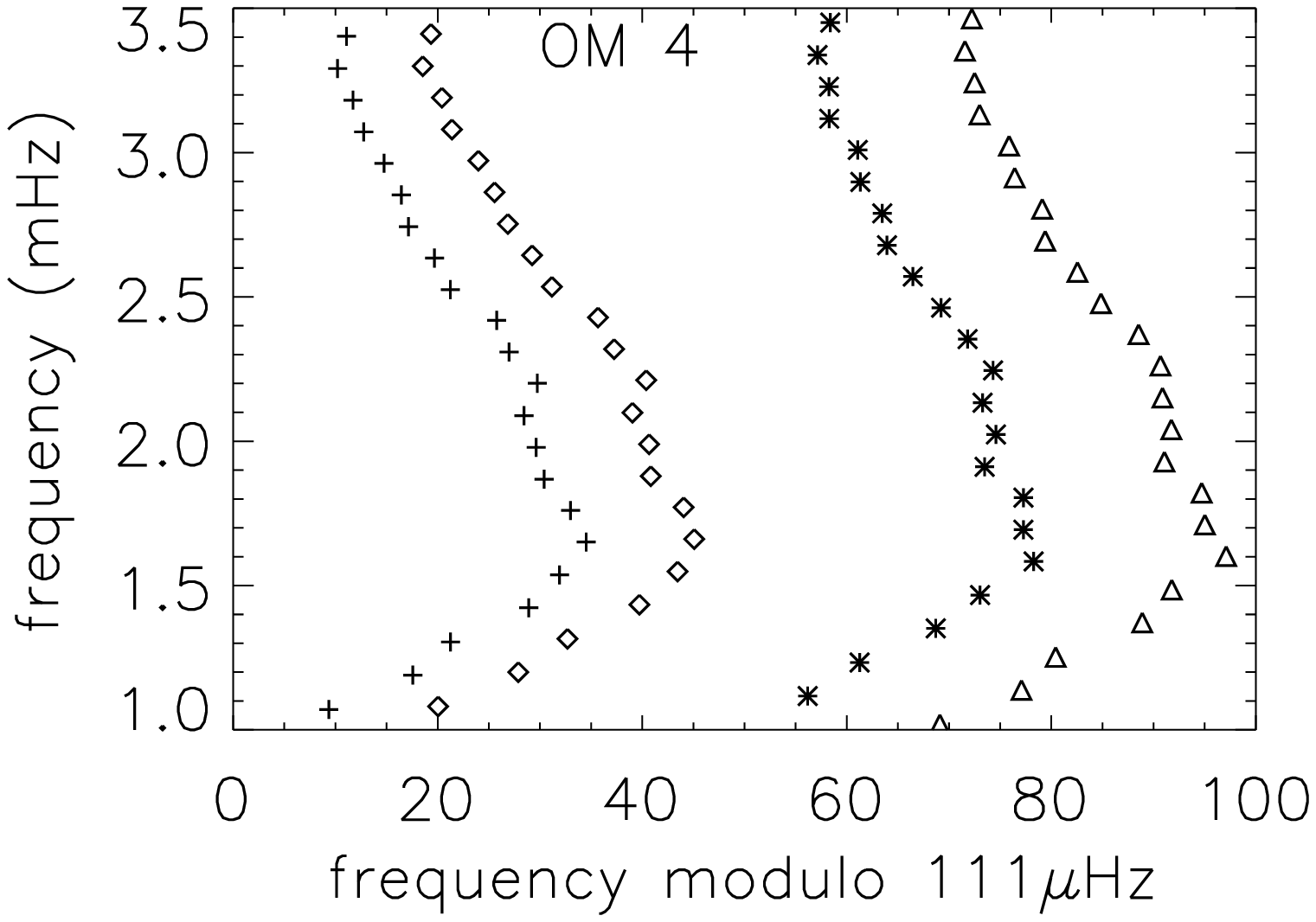}
\includegraphics[angle=0,totalheight=5.5cm,width=8cm]{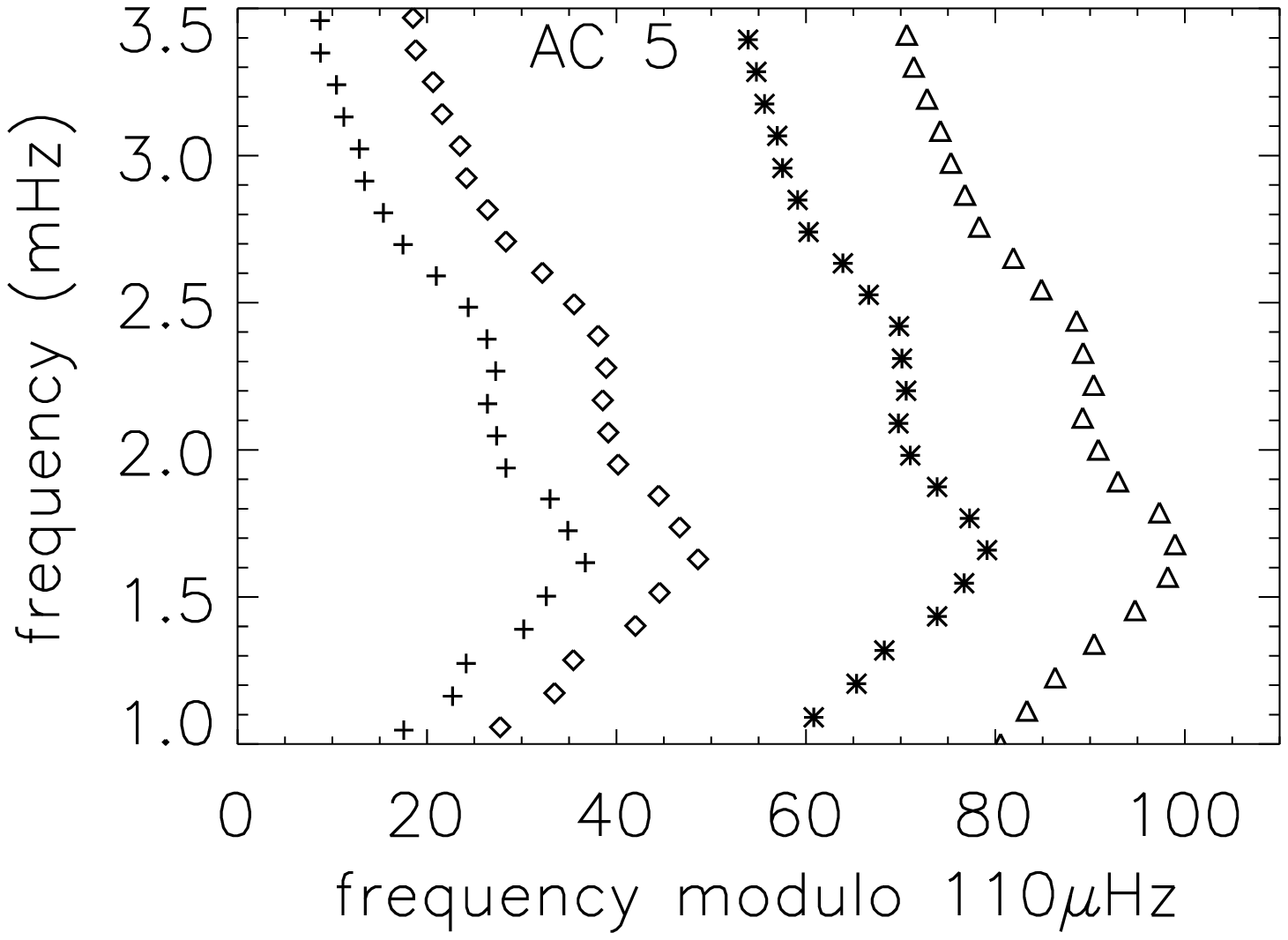}\includegraphics[angle=0,totalheight=5.5cm,width=8cm]{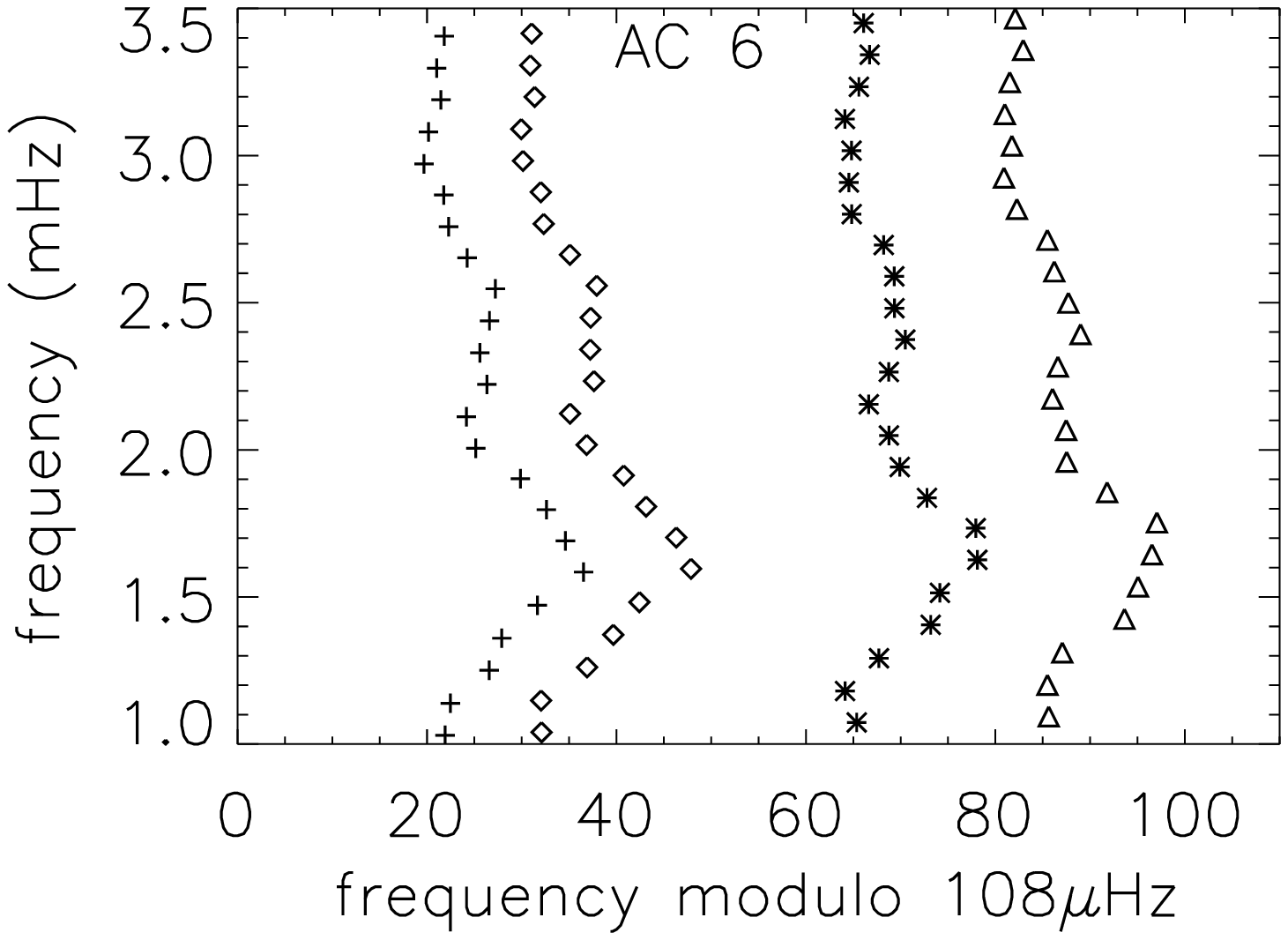}
\end{center}
\caption{Echelle diagrams for the overmetallic models 1 to 4 and the accretion models 5 and 6. Diamonds correspond to $\ell=0$, triangles to $\ell=1$, crosses to $\ell=2$ and asterisks to $\ell=3$. }
\label{fig3}
\end{figure*}

\begin{figure*}
\begin{center}
\includegraphics[angle=0,totalheight=5.5cm,width=8cm]{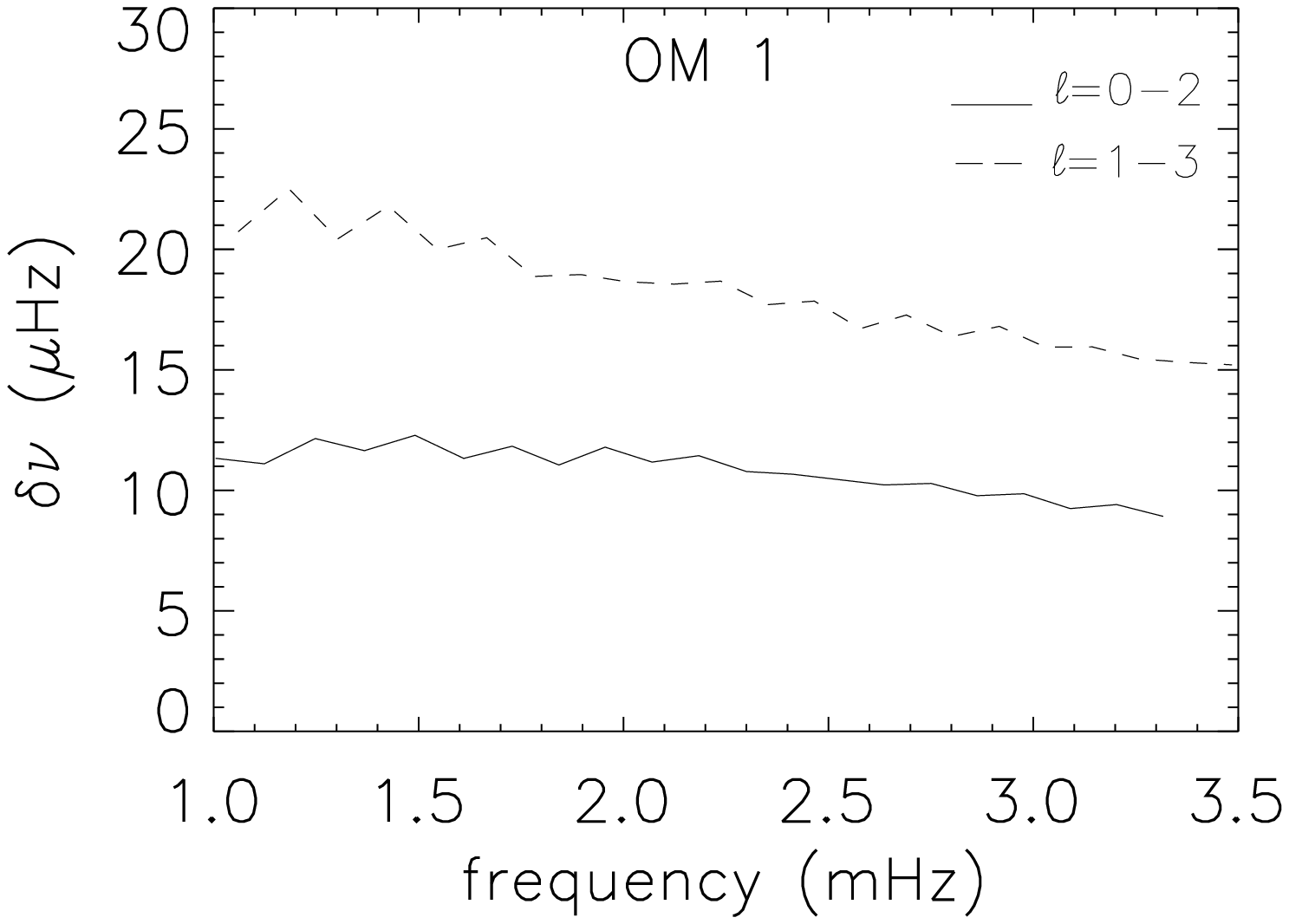}\includegraphics[angle=0,totalheight=5.5cm,width=8cm]{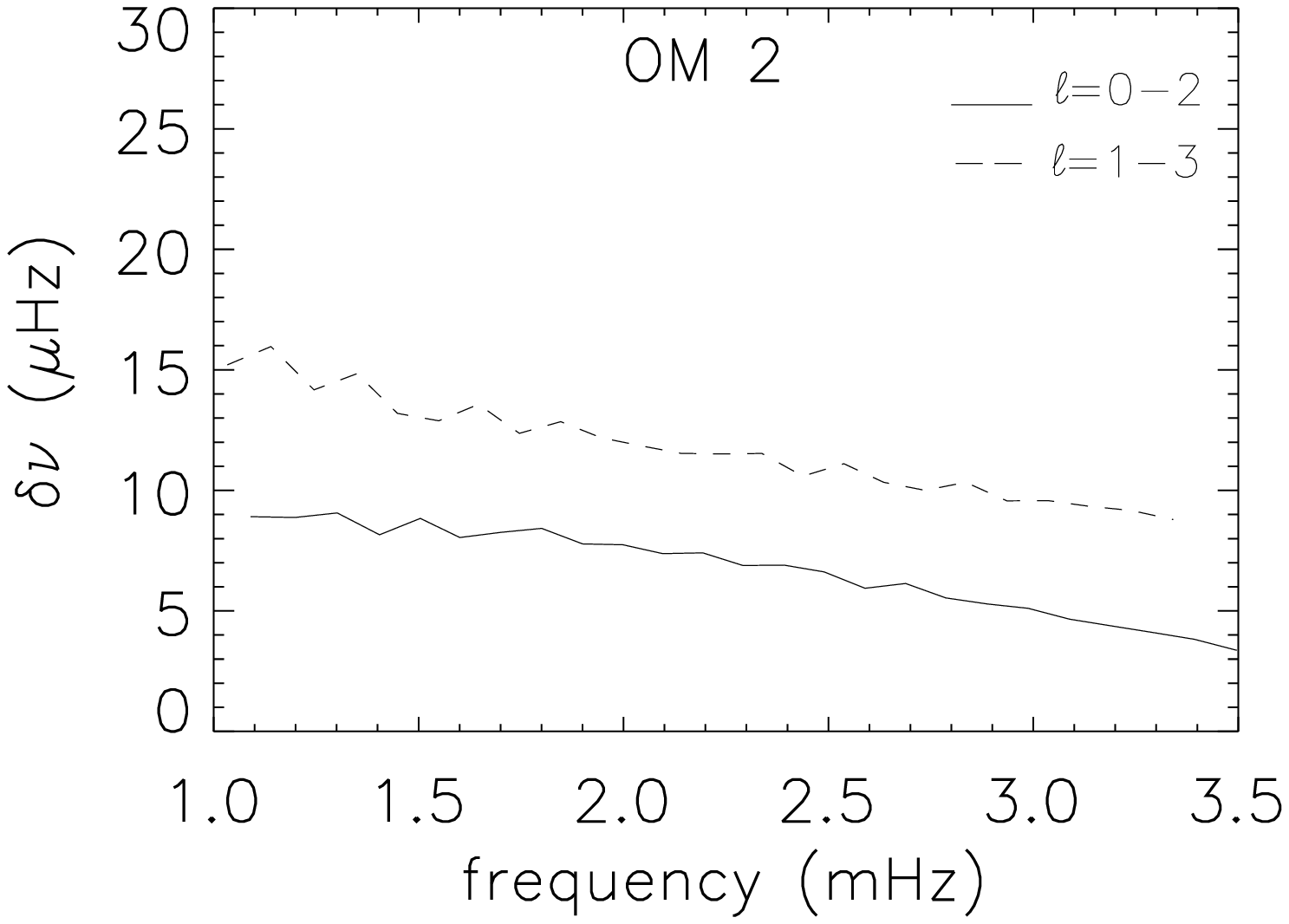}
\includegraphics[angle=0,totalheight=5.5cm,width=8cm]{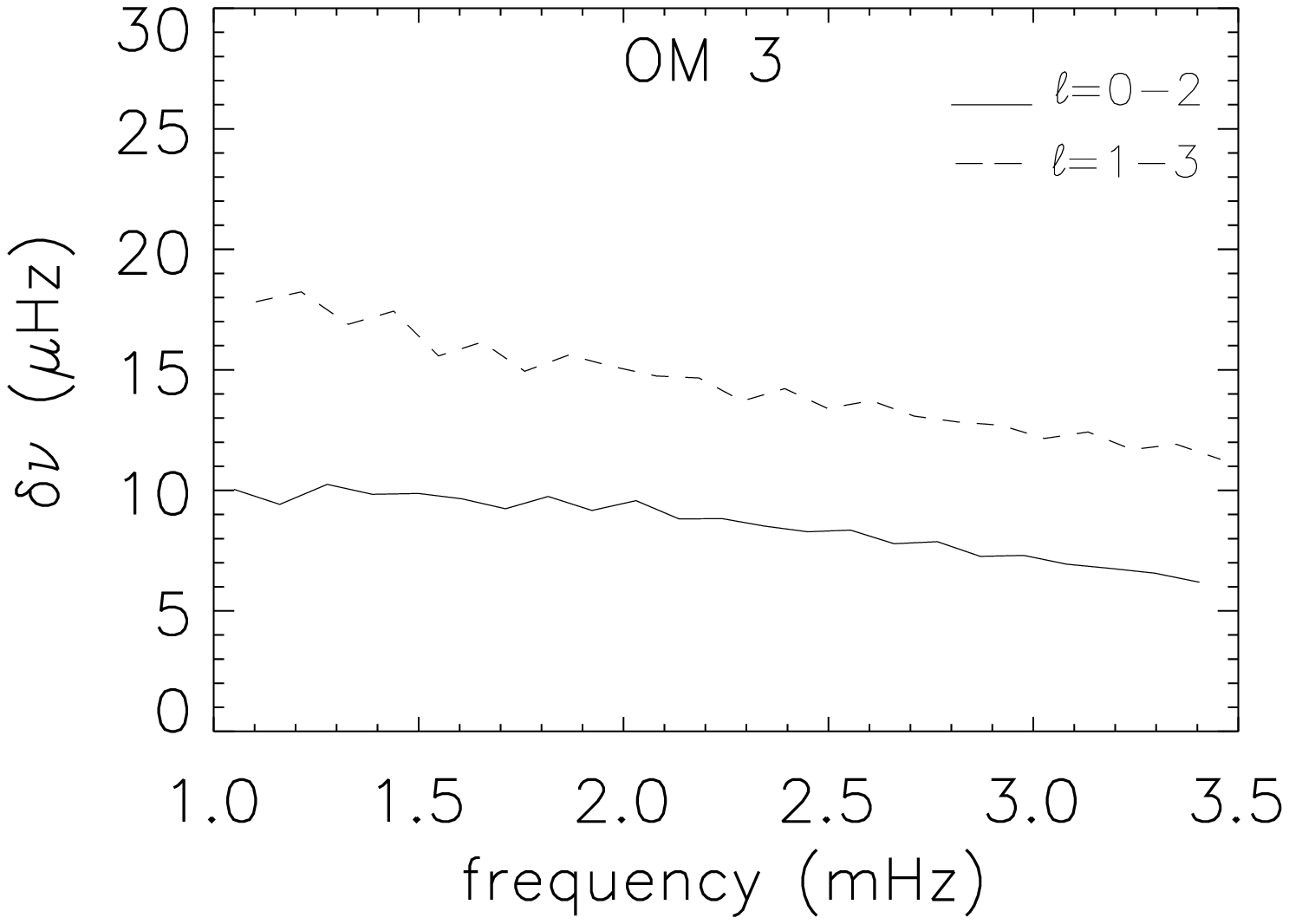}\includegraphics[angle=0,totalheight=5.5cm,width=8cm]{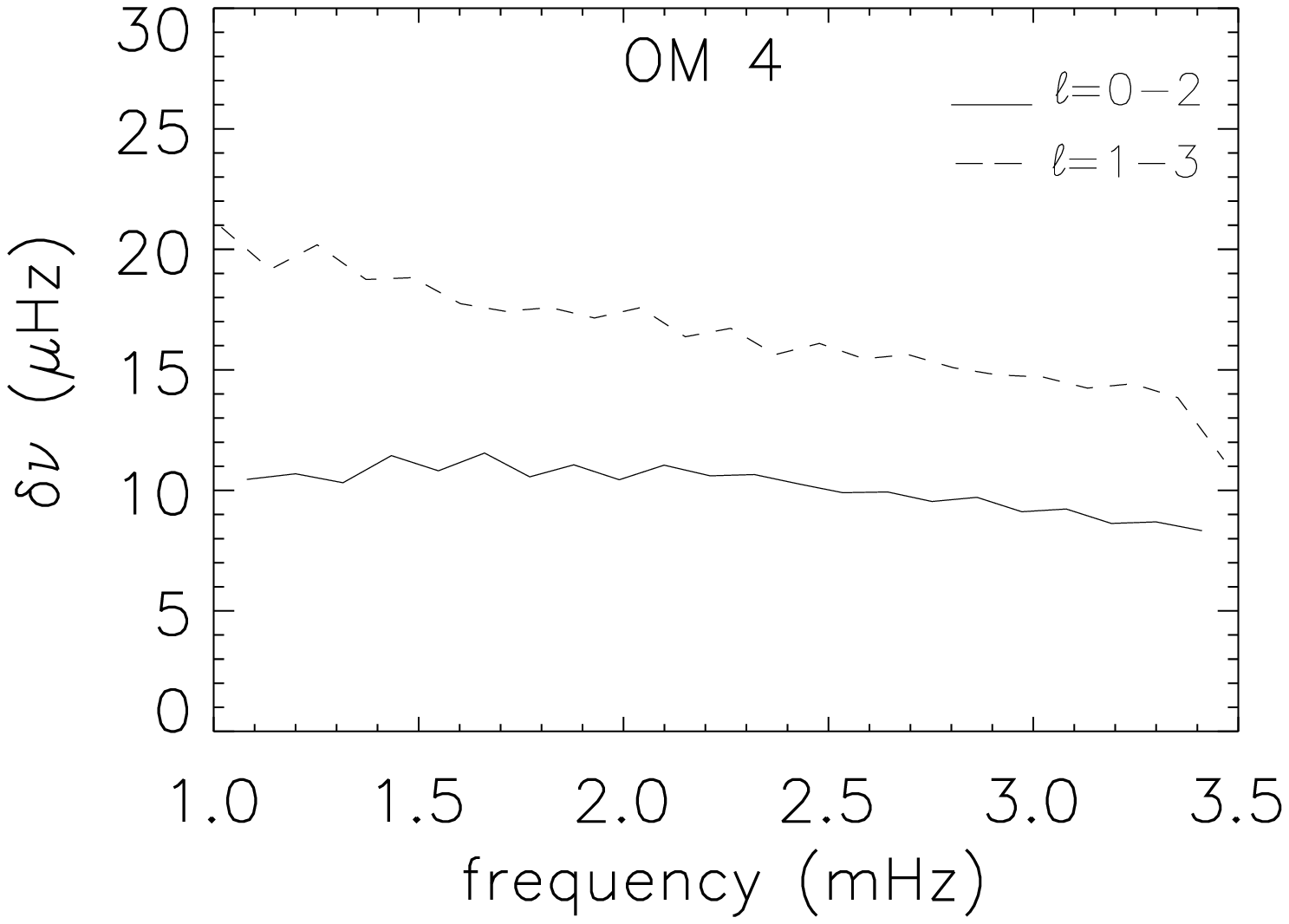}
\includegraphics[angle=0,totalheight=5.5cm,width=8cm]{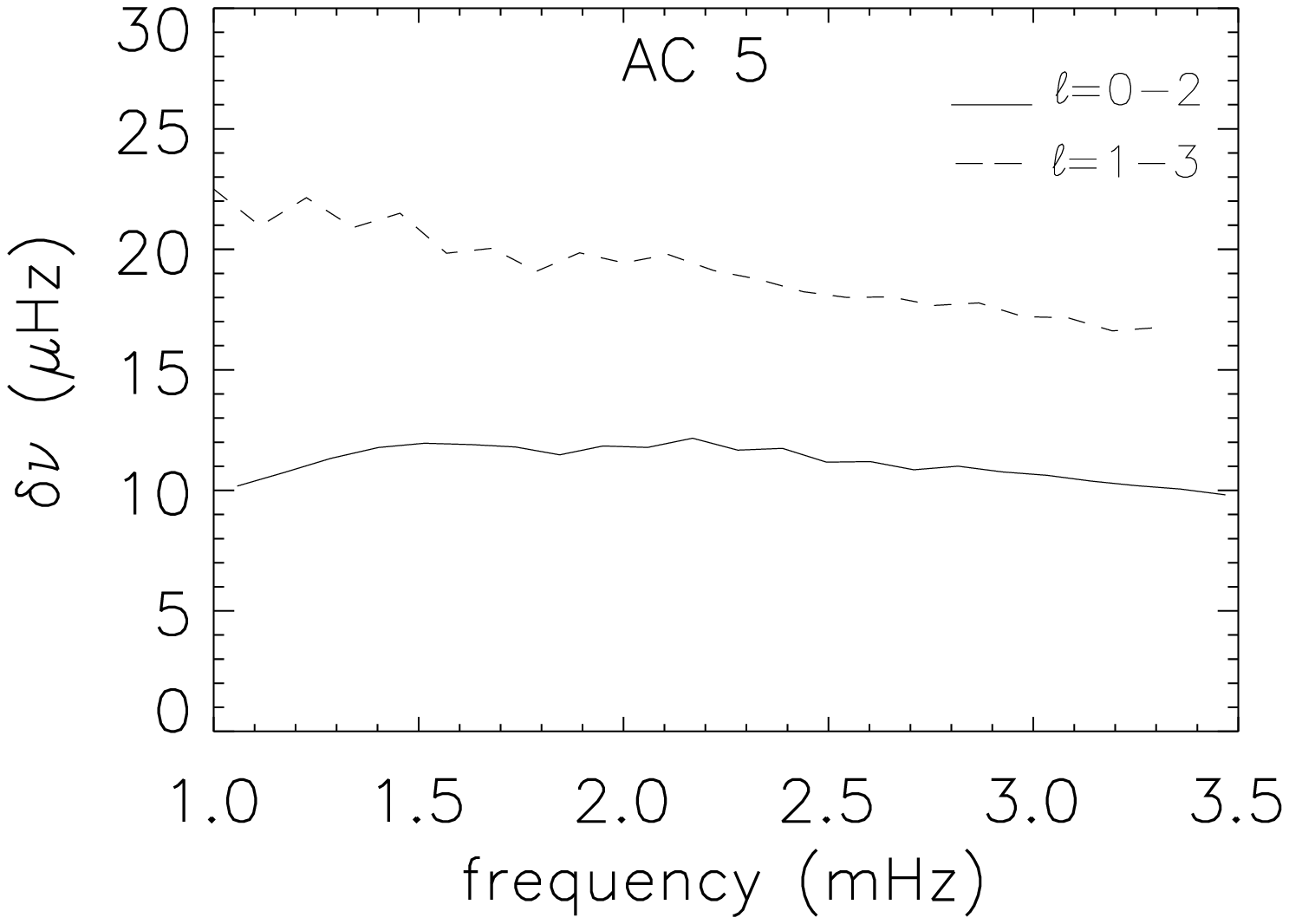}\includegraphics[angle=0,totalheight=5.5cm,width=8cm]{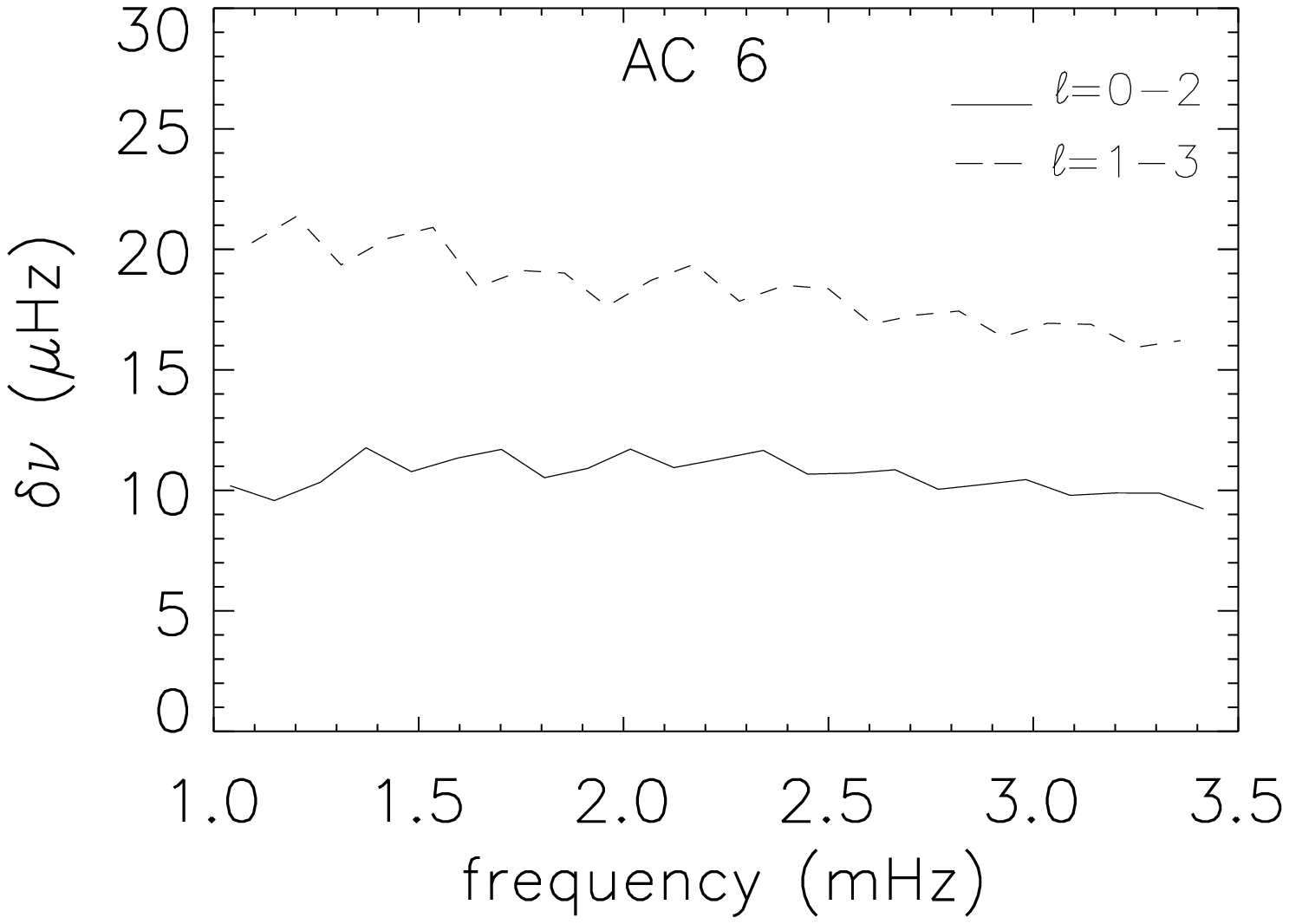}
\end{center}
\caption{ Small separations for the overmetallic models 1 to 4 and the accretion models 5 and 6. Solid lines are for the differences $\ell=0$ minus $\ell=2$ and dashed lines for the differences $\ell=1$ minus $\ell=3$.}
\label{fig4}
\end{figure*}

\begin{figure*}
\begin{center}
\includegraphics[angle=0,totalheight=5.5cm,width=8cm]{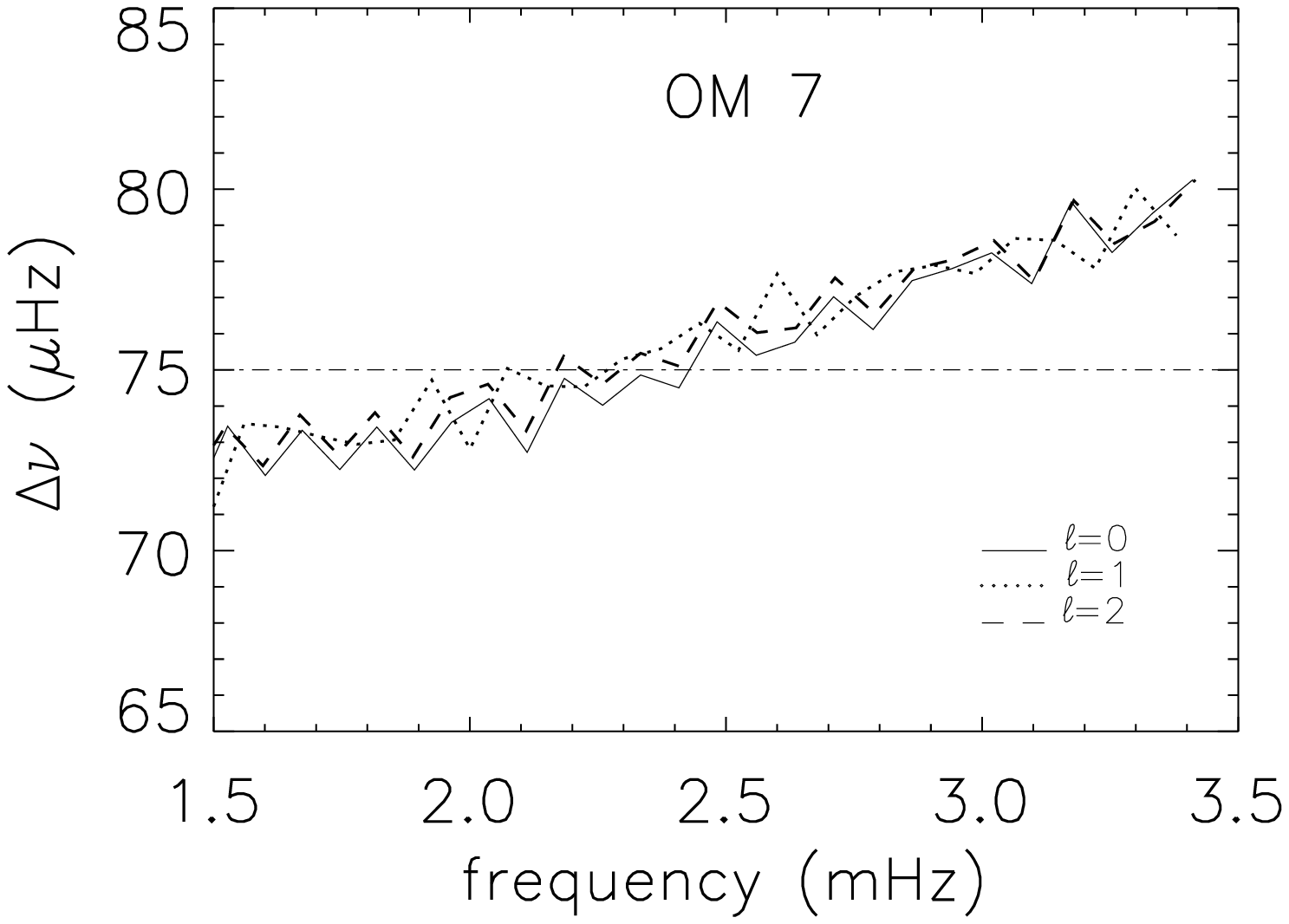}\includegraphics[angle=0,totalheight=5.5cm,width=8cm]{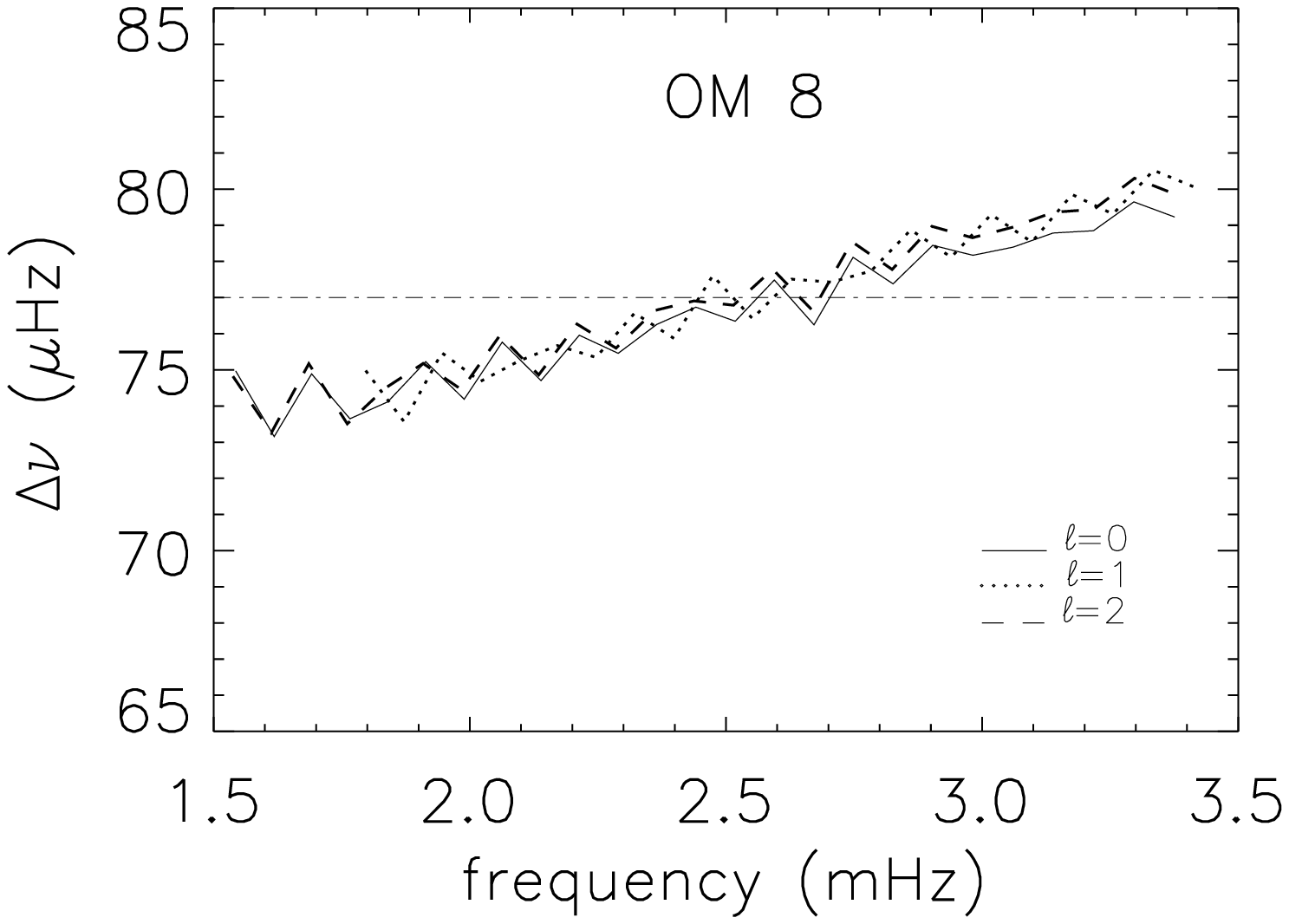}
\includegraphics[angle=0,totalheight=5.5cm,width=8cm]{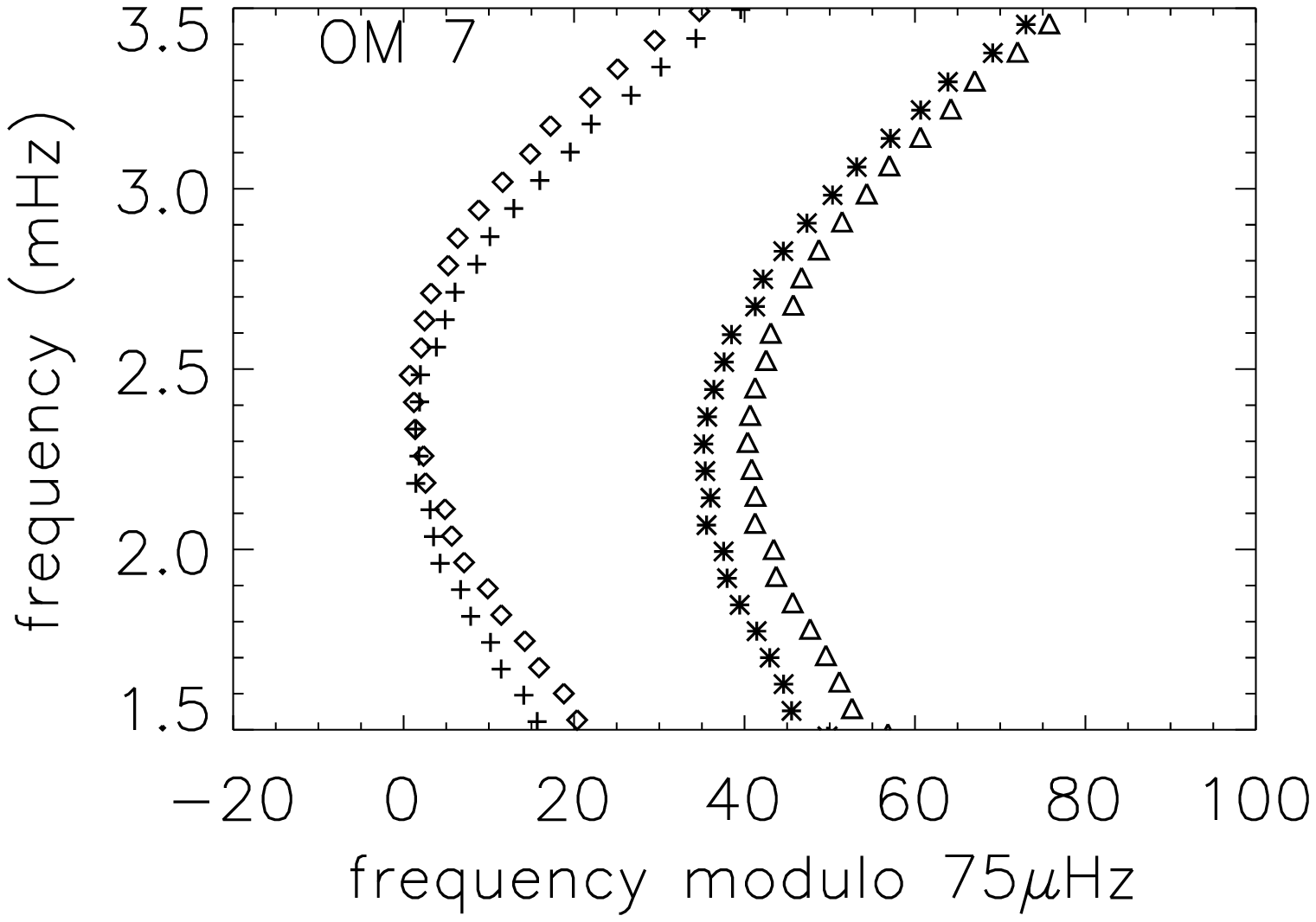}\includegraphics[angle=0,totalheight=5.5cm,width=8cm]{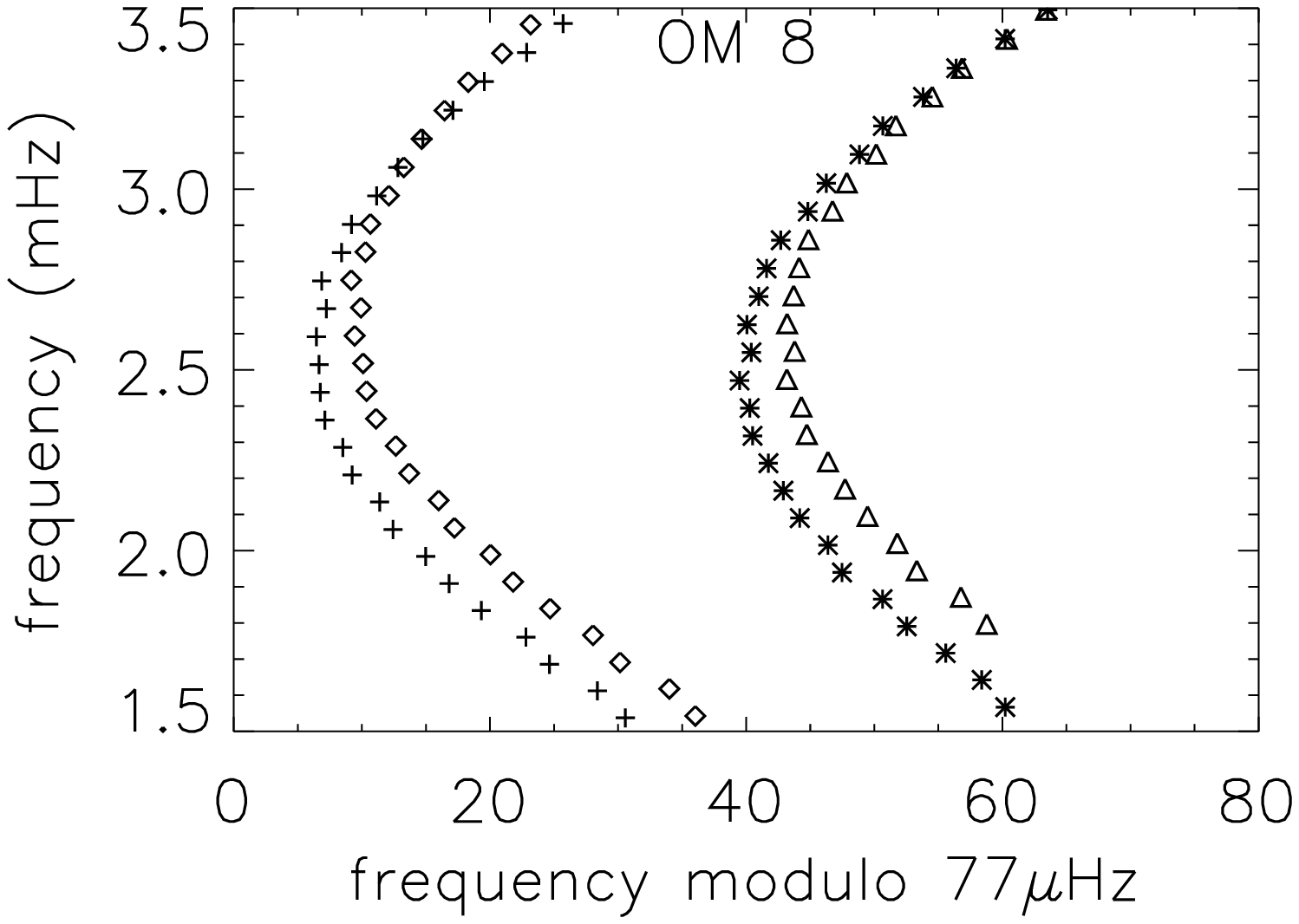}
\includegraphics[angle=0,totalheight=5.5cm,width=8cm]{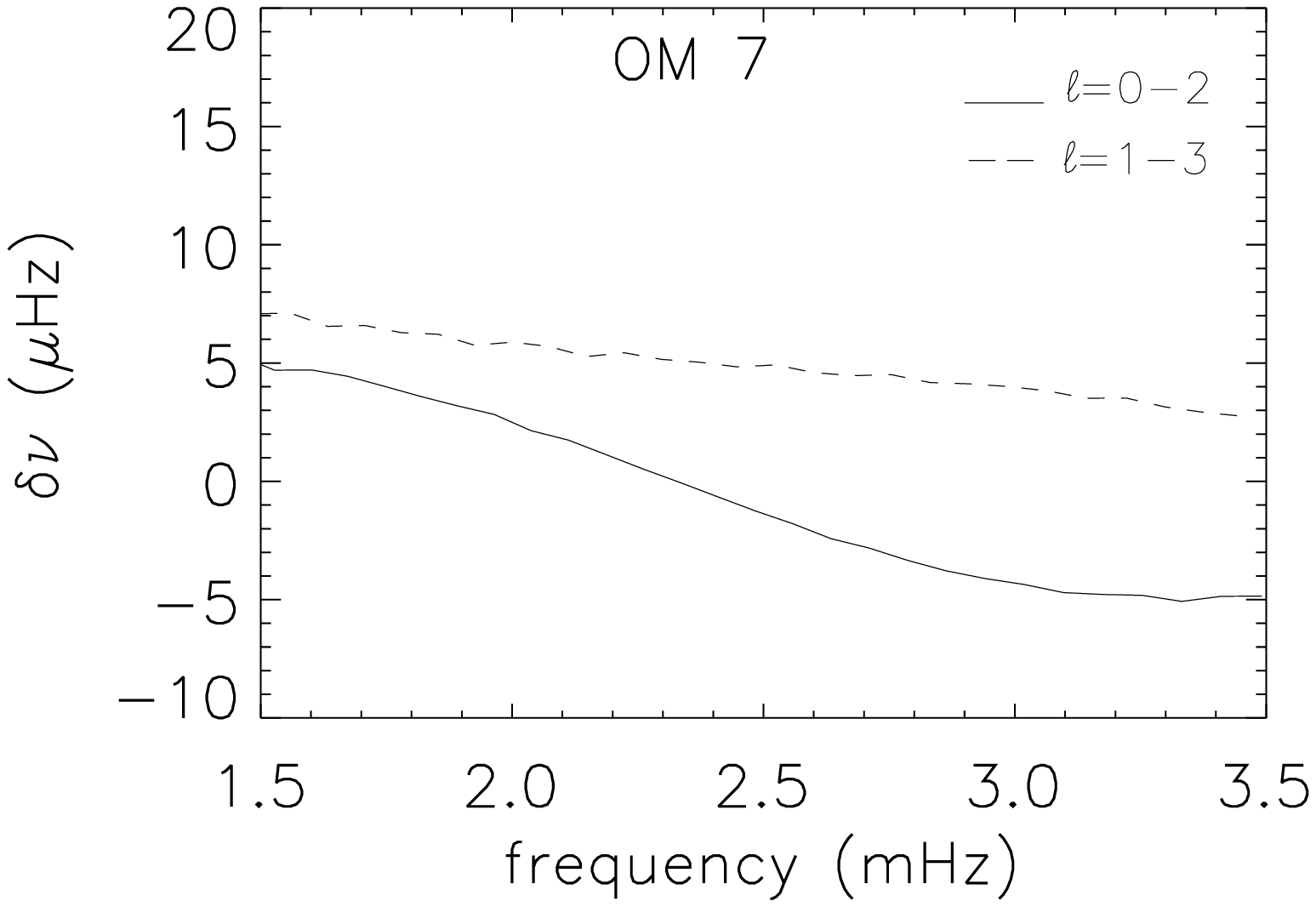}\includegraphics[angle=0,totalheight=5.5cm,width=8cm]{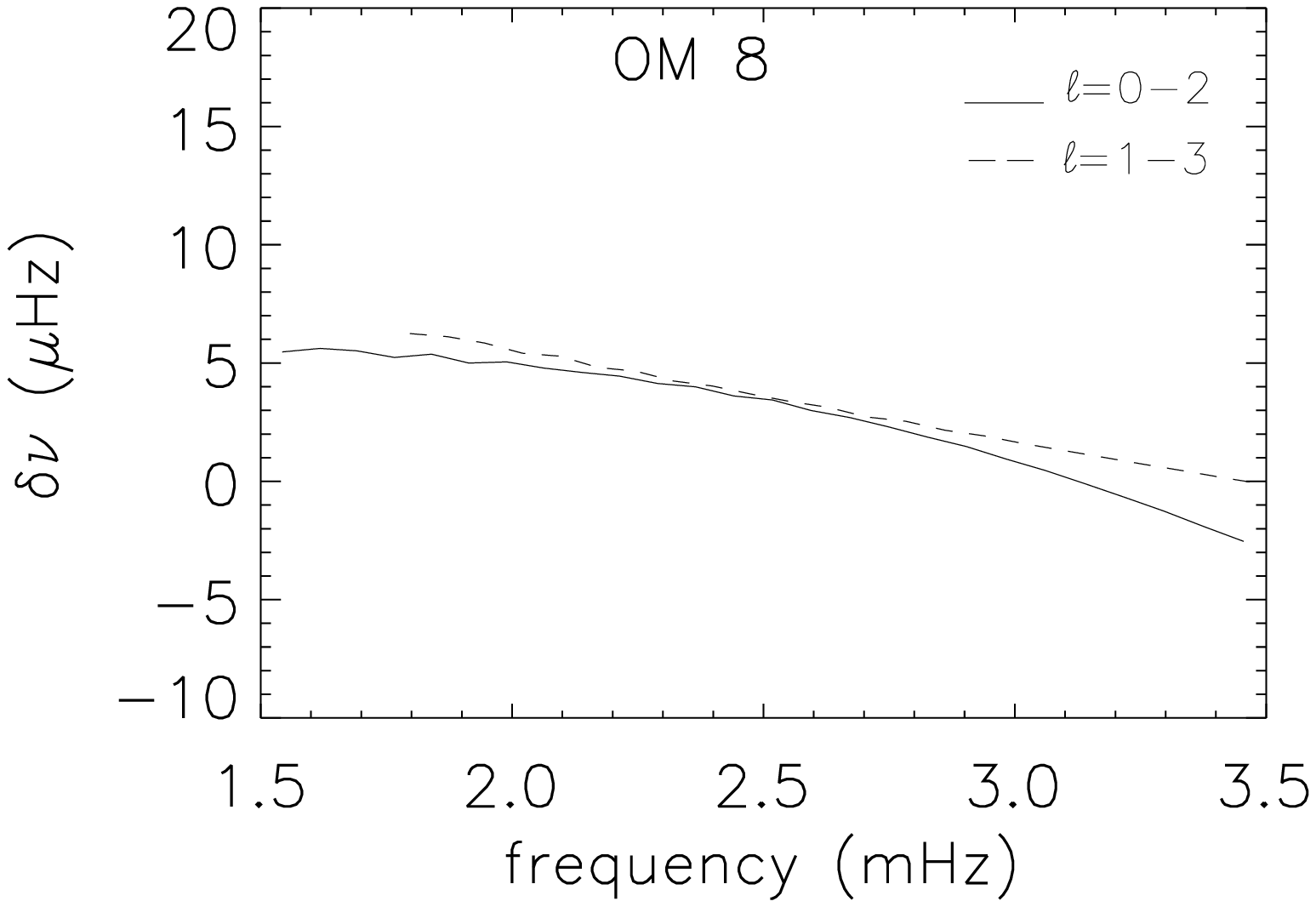}
\end{center}
\caption{ Large separations, echelle diagrams, and small separations for the overmetallic models 7 and 8. We note that, in the echelle diagrams, the $\ell=0$ and $\ell=2$ lines cross at frequencies of 2.3 mHz (model 7) and 3.15 mHz (model 8) while the $\ell=1$ and $\ell=3$ lines becomes very close for large frequencies. This is due to their helium-rich cores (see text).}
\label{fig5}
\end{figure*}

\begin{figure*}
\begin{center}
\includegraphics[angle=0,totalheight=5.5cm,width=8cm]{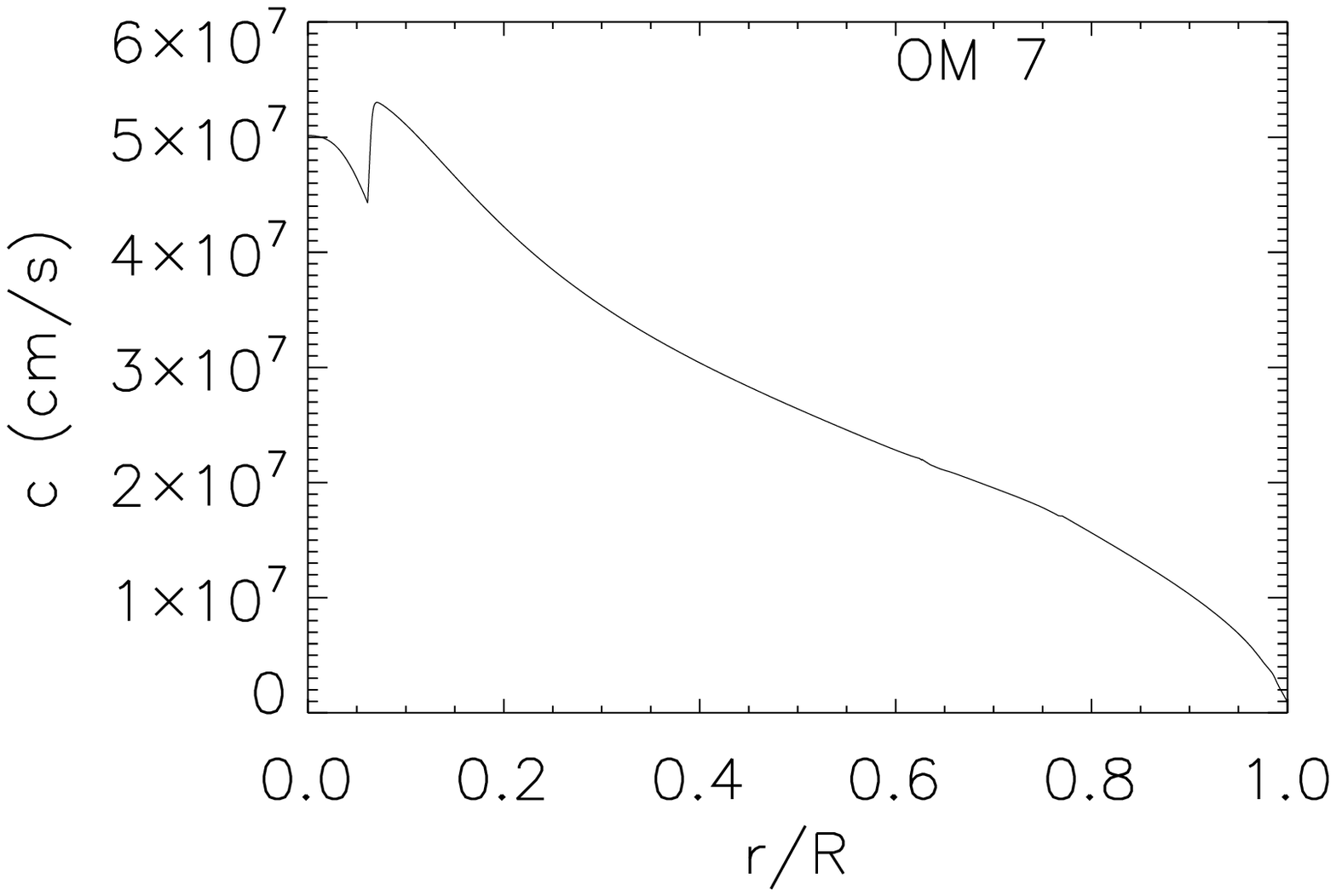}\includegraphics[angle=0,totalheight=5.5cm,width=8cm]{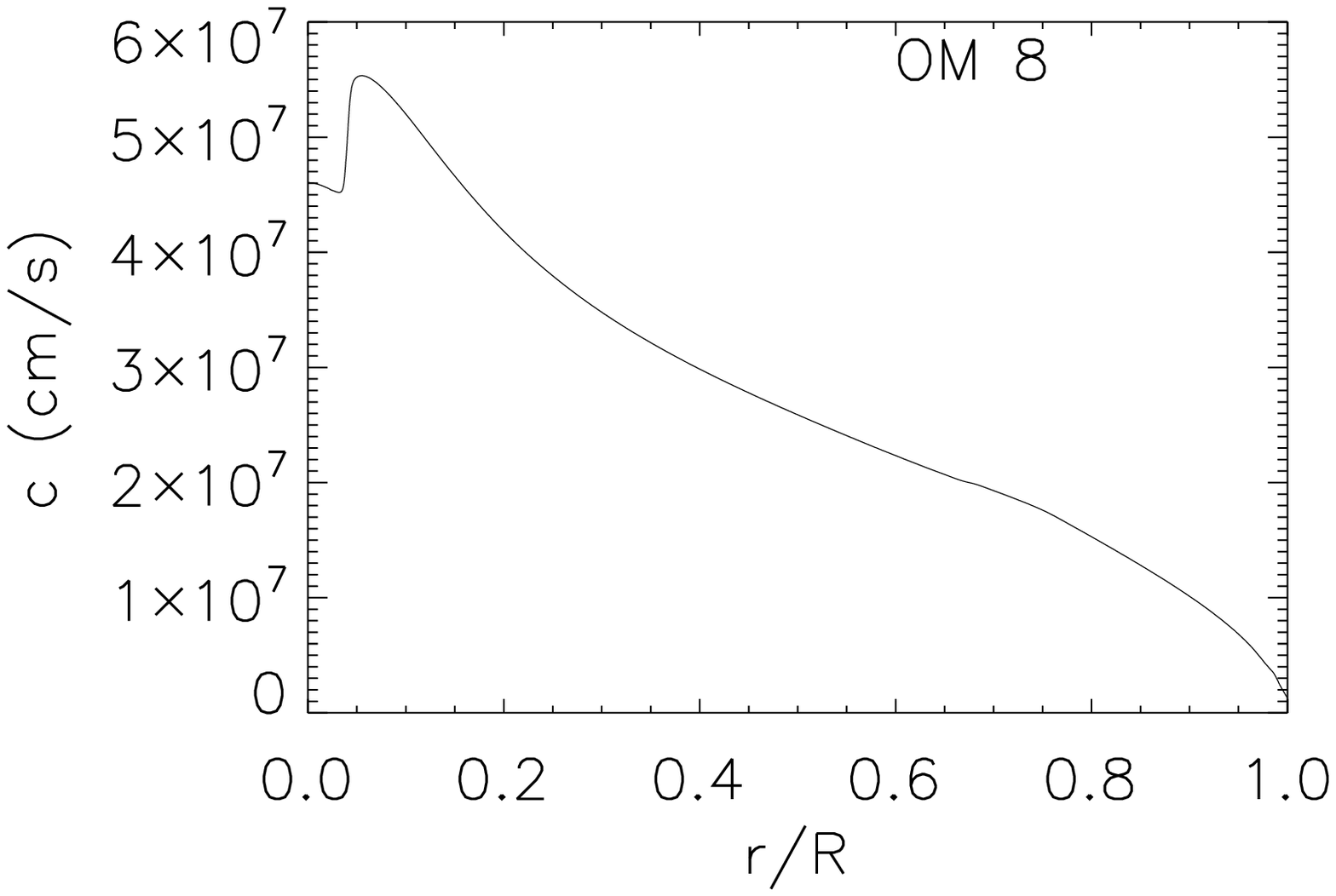}
\includegraphics[angle=0,totalheight=5.5cm,width=8cm]{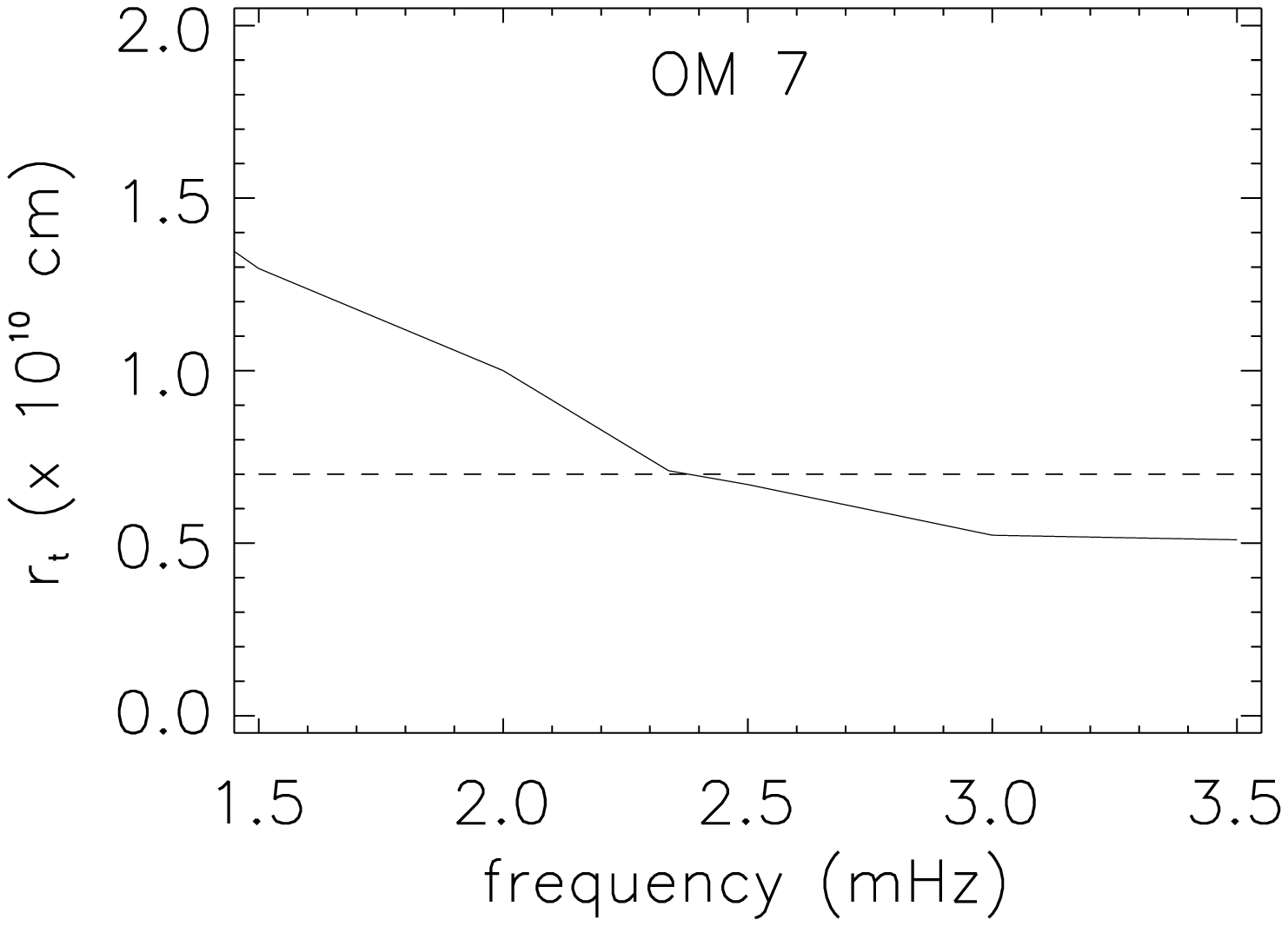}\includegraphics[angle=0,totalheight=5.5cm,width=8cm]{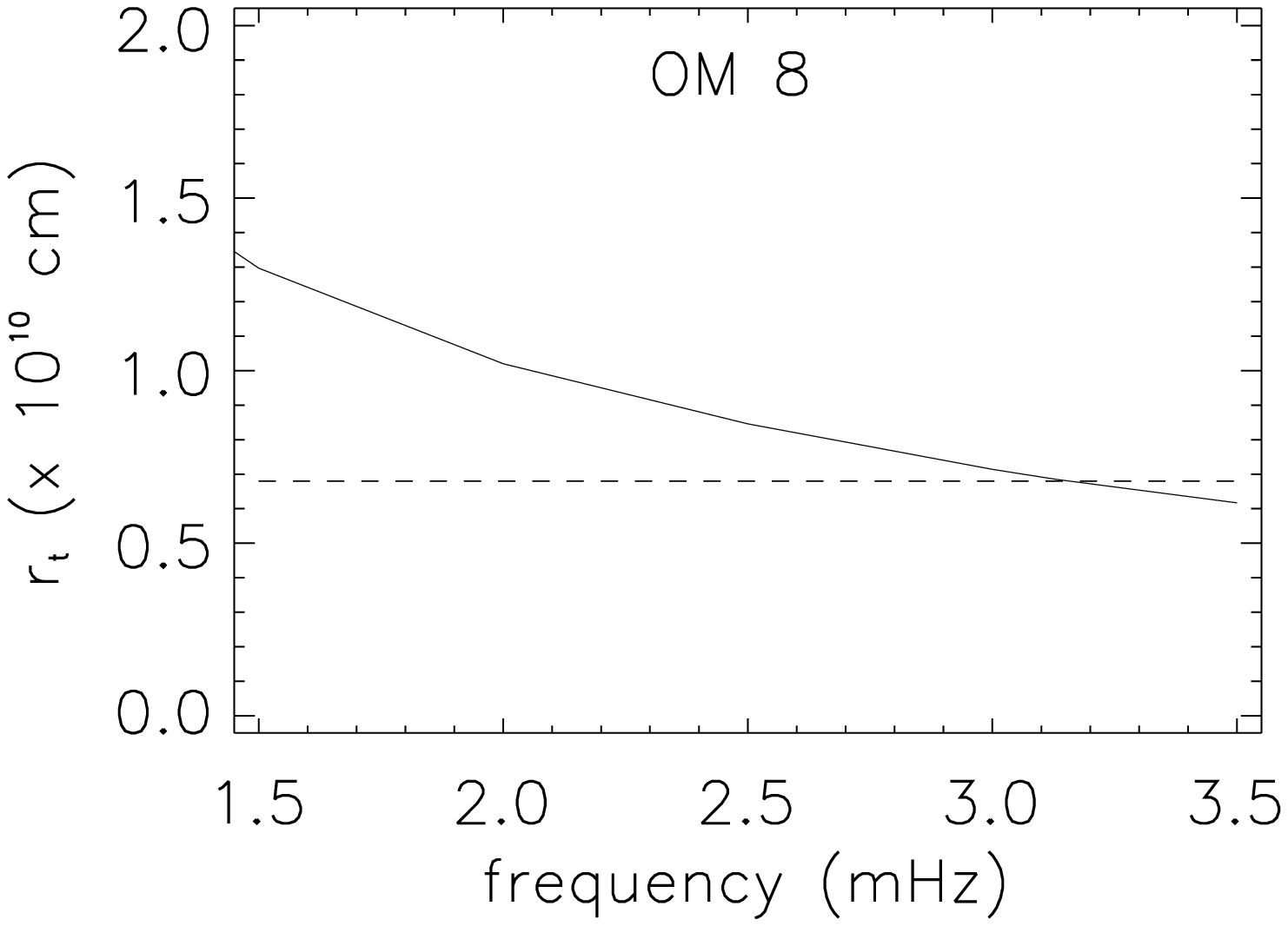}
\end{center}
\caption{ Upper panels: sound speed profiles for the overmetallic models 7 and 8. The sound velocities present a sharp variation at the boundary of the helium-rich core, which is convective in model 7 but radiative in model 8. Lower panels: turning point $r_t$ of the $\ell=2$ waves, computed as a function of the frequency of the waves (solid lines). The dashed lines represent the radius of the helium-rich core. We see that the turning point reaches the core boundary at the same frequency as that of the crossing point in the echelle diagrams.}
\label{fig6}
\end{figure*}

\begin{figure*}
\begin{center}
\includegraphics[angle=0,totalheight=5.5cm,width=8cm]{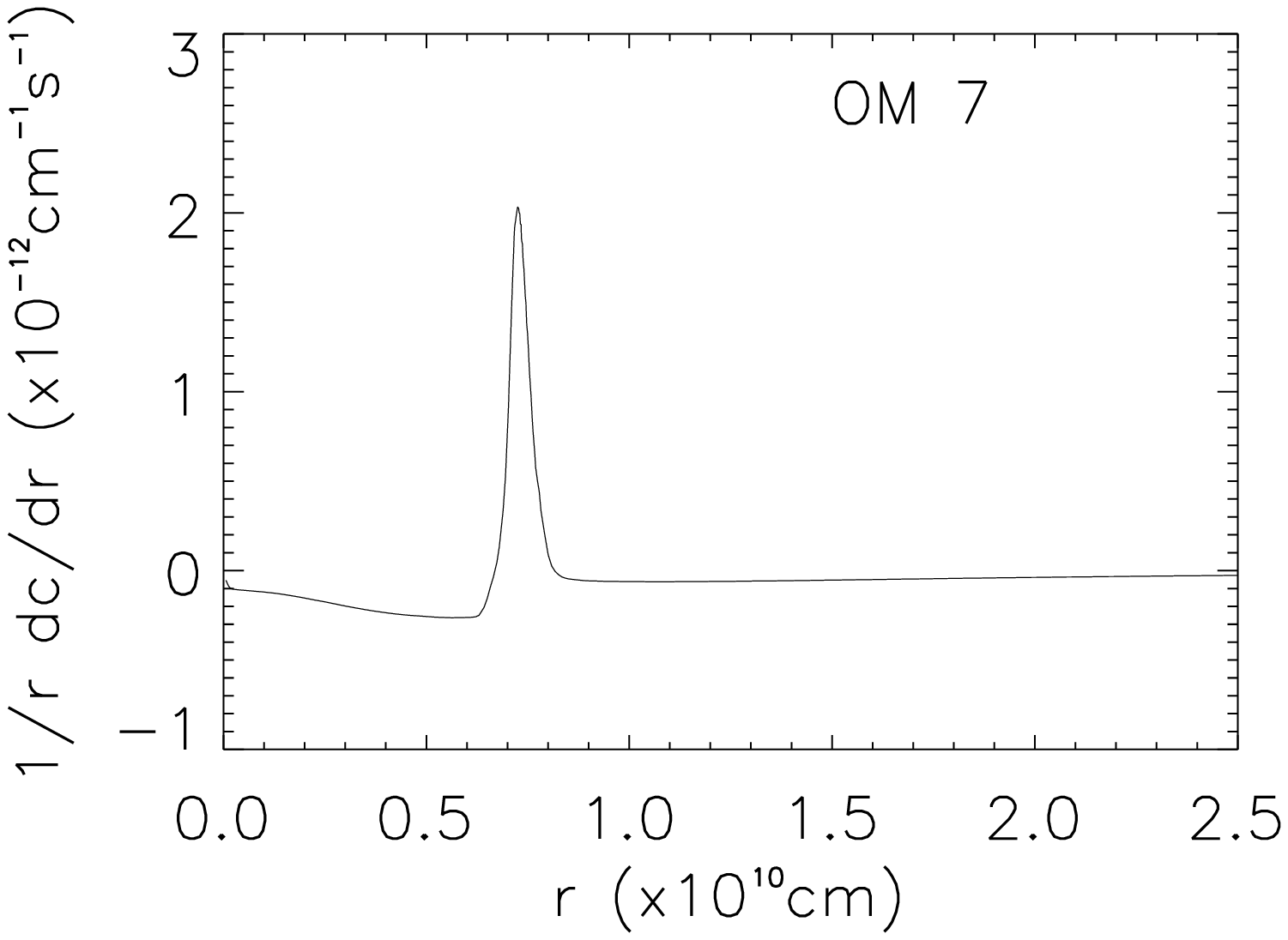}\includegraphics[angle=0,totalheight=5.5cm,width=8cm]{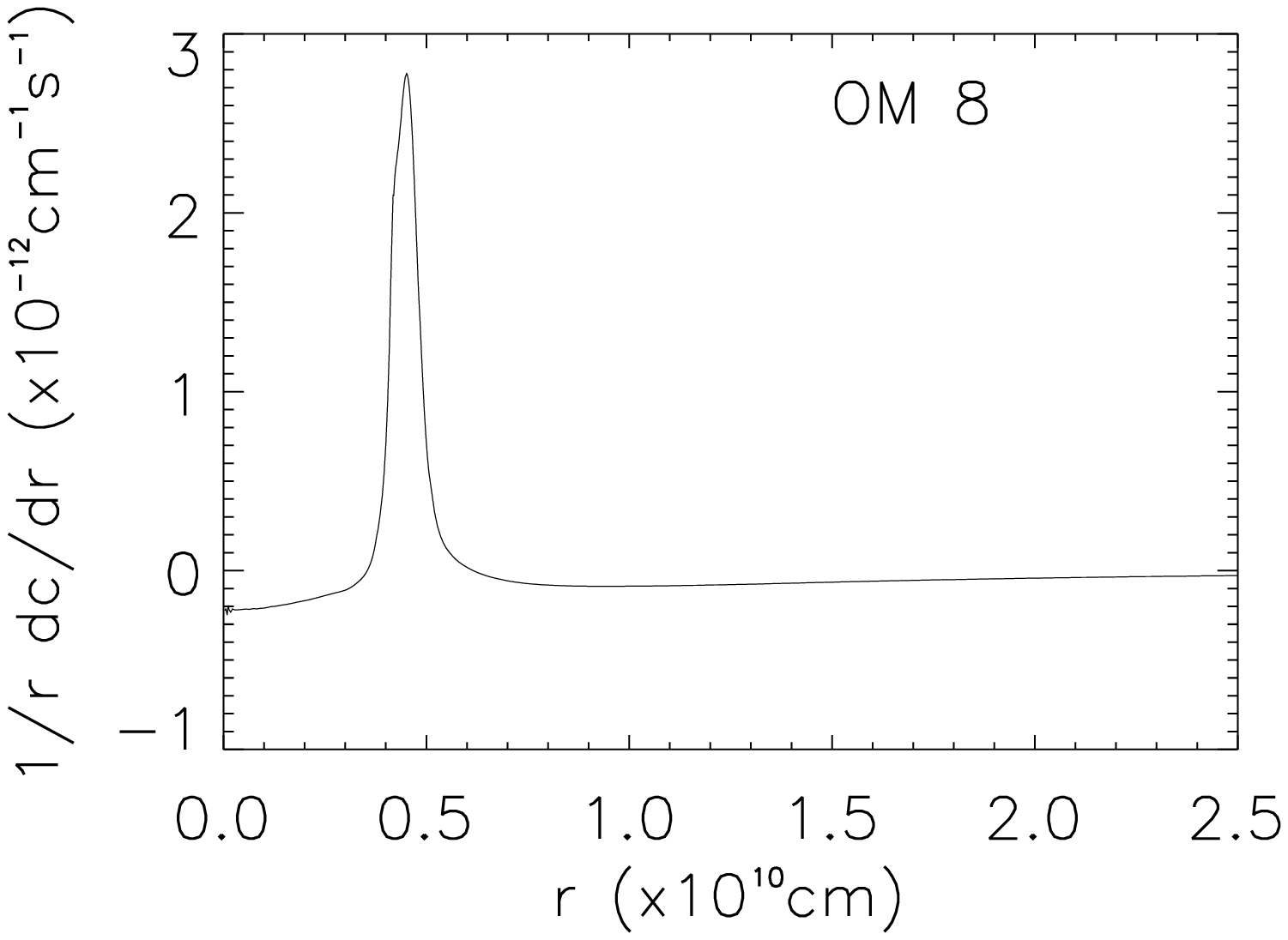}
\includegraphics[angle=0,totalheight=5.5cm,width=8cm]{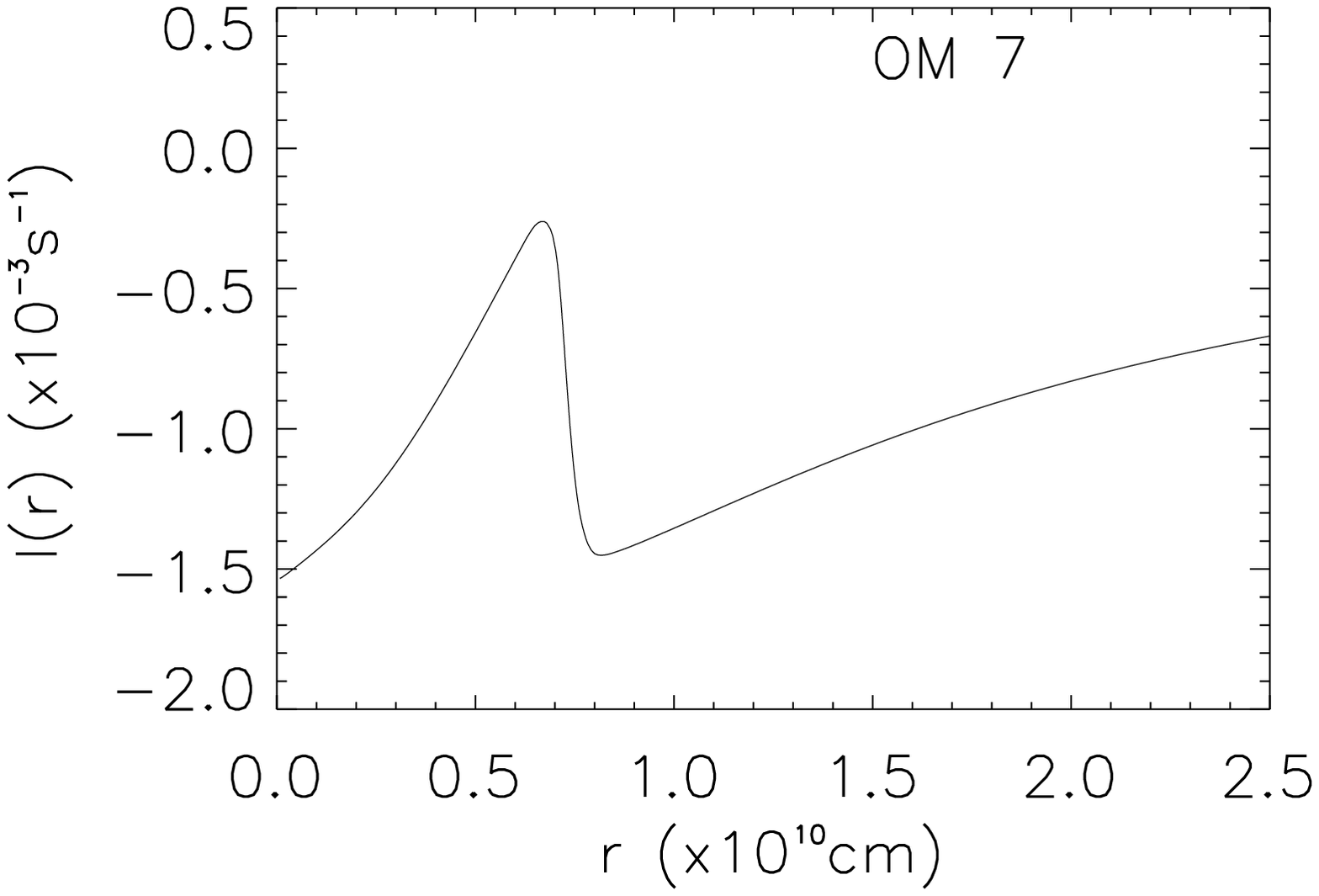}\includegraphics[angle=0,totalheight=5.5cm,width=8cm]{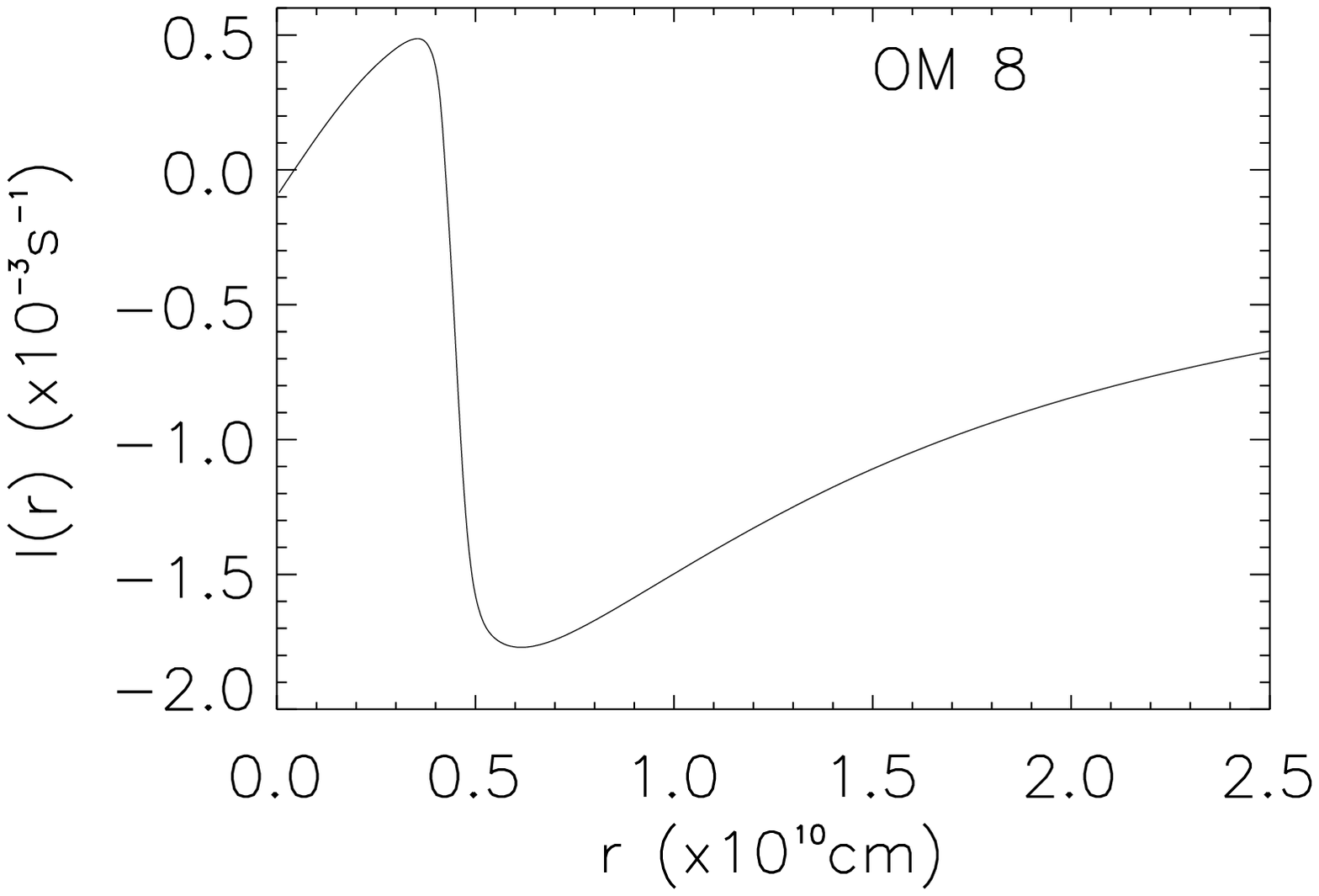}
\end{center}
\caption{ Upper panels: variations of the product $\frac{1}{r} \frac{dc}{dr}$ in the central regions of the stars. Lower panels: integral $I(r)=\int_{r_{t}}^{R}\frac{1}{r}\frac{dc}{dr}dr$}
\label{fig7}
\end{figure*}

\subsection{Evolutionary tracks for overmetallic models}

We have first computed evolutionary tracks for overmetallic models, using the three different values of the metallicity as given by the groups of observers: [Fe/H] = 0.19 (TOSKS05 and FV05), 0.23 (SIM04 and GM06) and 0.27 (GLTR01). The results obtained in the log~$g$ - log~$T_{eff}$ diagram are displayed in Fig.~\ref{fig1} (upper and middle panels). Each graph corresponds to one value of the metallicity. The five observational error boxes are shown in each graph but those which correspond to the same metallicity as the models are drawn in thicker lines. The only acceptable models computed with a given metallicity are those which cross the error boxes corresponding to the same metallicity as derived by the observers. Two different graphs are presented for the [Fe/H] = 0.19 case, because the TOSKS05 box can be crossed in two different ways. On the graph presented on the upper left, the evolutionary tracks cross this box at the end of the main sequence, while on the graph presented on the upper right, this happens at the beginning of the subgiant phase. Of course the masses involved are quite different as well as the internal structure of the stars, as will be seen below.

\subsection{Evolutionary tracks for accretion models}
 
The lower panels of Fig.~\ref{fig1} shows the evolutionary tracks of models with accretion for two metallicities : [Fe/H] = 0.23 and [Fe/H] = 0.27. We observe that, for the same value of [Fe/H], the accretion models which cross the error boxes are more massive and younger than the overmetallic ones. We also note that for all the considered cases, except the special one for which the star should be a subgiant, the range of stellar masses is from 1.18$M_{\odot}$ to 1.25$M_{\odot}$ so that all the models (overmetallic and accretion) develop convective cores during evolution.

\section{Models and seismic tests}

We have computed a large number of models along the evolutionary tracks with the constraints of having consistent metallicities, effective temperatures and gravities as given by each group of observers. We choose eight of these models for further asteroseismic studies.

\subsection{Choice of models}

The characteristics of the models chosen for asteroseismic studies are displayed in Table~\ref{tab2} and Table~\ref{tab3}. The overmetallic models labelled 1 to 4 (OM) correspond to main-sequence stars. Their seismic analysis is given in Sect.~3.2. Models computed with the accretion hypothesis, labelled 5 and 6 (AC) are discussed in Sect.~3.3. The evolutionary tracks which cross the error box of Takeda et al. (\cite{takeda05}) lead to quite different models, either at the end of the main-sequence or at the beginning of the subgiant branch. We choose to discuss two of them, labelled 7 and 8 (OM) in a special section (Sect~4). These models, which have large luminosities compared to those derived from the Hipparcos parallax, present specially interesting characteristics: in both cases the small separations become negative at some particular frequency. Note that in model 8 (OM) the convective core has disappeared while it did develop along the evolutionary track. For this reason, model 8 (OM) possesses a radiative helium core with a sharp boundary, while model 7 still has a convective core. These interesting models and the underlying physics are discussed in Sect.~4.

\begin{table*}
\caption{External parameters of eight stellar models which could account for HD~52265 within the present spectroscopic constraints: mass, age, surface gravity, effective temperature, luminosity, surface metallicity, the average large separations computed for these models are given in column 8.}
\label{tab2}
\begin{flushleft}
\begin{tabular}{cccccccc} \hline
\hline
Model & M$_{\star}$ (\msol) & Age (Gyr) & log $g$ & log $T_eff$ (K) & log L/\lsol &[Fe/H] & $\Delta \nu$ ($\mu$Hz) \cr
  \hline \hline
 1 (OM) & 1.21  & 1.275 & 4.360 & 3.7915 & 0.268 & 0.23 & 115\cr 
 2 (OM) & 1.18  & 3.427 & 4.293 & 3.7839 & 0.318 & 0.19 & 100\cr
 3 (OM) & 1.19  & 2.530 & 4.320 & 3.7867 & 0.288 & 0.23 & 107\cr
 4 (OM) & 1.22  & 1.544 & 4.300 & 3.7910 & 0.291 & 0.27 & 111\cr
  \hline
 5 (AC) & 1.22  & 0.390 & 4.300 & 3.7892 & 0.290 & 0.23 & 110\cr
 6 (AC) & 1.23  & 0.603 & 4.320 & 3.7892 & 0.319 & 0.27 & 108\cr
  \hline
 7 (OM) & 1.31  & 3.219 & 4.125 & 3.7825 & 0.519 & 0.19 & 75\cr
 8 (OM) & 1.20  & 4.647 & 4.125 & 3.7831 & 0.491 & 0.19 & 77\cr
\hline
\end{tabular}
\end{flushleft}
\end{table*}

\begin{table*}
\caption{Internal parameters for the same models as given in Table~2: radius,  fractional radius and mass of the convective core, fractional radius of the outer convective zone, initial helium and metal mass fraction, present surface helium and metal mass fraction, and present central helium mass fraction; for model 8, which has no more convective core, columns 3 and 4 present the radius and mass fractions of the helium core.}
\label{tab3}
\begin{flushleft}
\begin{tabular}{cccccccccc}\hline
\hline
Model & R$_{\star}$ (cm) & r$_{cc}$/R$_{\star}$ & M$_{cc}$/M$_{\star}$ & r$_{ec}$/R$_{\star}$ & Y$_0$ & Z$_0$ & Y$_s$ & Z$_s$ & Y$_c$\cr
\hline \hline
 1 (OM) & 8.266e10 & 0.0501 & 0.0121 & 0.798 & 0.2930 &0.0282 & 0.270 & 0.0269  & 0.46\cr
 2 (OM) & 9.014e10 & 0.0559 & 0.0032 & 0.762 & 0.2879 & 0.0260 & 0.2408 & 0.0236 & 0.728 \cr
 3 (OM) & 8.643e10 & 0.0515 & 0.019 & 0.776 & 0.2930 &0.0282 & 0.2545 & 0.0261 & 0.624 \cr
 4 (OM) & 8.510e10 & 0.0573 & 0.020 & 0.793 & 0.2984 &0.0306 & 0.2695 & 0.0288 & 0.501\cr
\hline
 5 (AC) & 8.570e10 & 0.0433 & 0.0077 & 0.82 & 0.2714 & 0.0189 & 0.2536 & 0.0808 & 0.3244\cr
 6 (AC) & 8.864e10 & 0.0412 & 0.0086 & 0.83 & 0.2714 & 0.0189 & 0.2343 & 0.1094 & 0.3581\cr
\hline
 7 (OM) & 11.51e10 & 0.0610 & 0.090 & 0.767 & 0.2879 & 0.0260 & 0.2397 & 0.0238 & 0.876\cr
 8 (OM) & 11.11e10 & (0.0612) & (0.1077) & 0.764 & 0.2879 & 0.0260 & 0.2405 & 0.0238 & 0.971\cr
\hline
\end{tabular}
\end{flushleft}
\end{table*}

\subsection{Seismic analysis of models 1 to 4: large separations, small separations and echelle diagrams}

Adiabatic oscillation frequencies were computed for the models chosen in Sect.~2 using the Brassard et al. (\cite{brassard92}) code, as in Bazot \& Vauclair (\cite{bazot04}). The frequencies were computed for angular degrees $\ell=0$ to $\ell=3$ and radial orders ranging typically from 4 to 100.
The corresponding echelle diagrams, where the frequencies are plotted in ordinates and the same frequencies modulo the large separation are plotted in abscissae, are given in Fig.~\ref{fig3}. They present characteristic differences large enough to be detectable with the CoRoT satellite, which should reach a frequency resolution of the order of 0.1 $\mu$Hz for the long observational runs of 150 days. 

We also computed the small separations, $\delta \nu = \nu_{n,l} - \nu_{n-1,l+2}$, which are very sensitive to the deep stellar interior (Tassoul \cite{tassoul80}, Roxburgh \& Vorontsov \cite{roxburgh94}, Gough \cite{gough86}). They are presented in Fig.~\ref{fig4}. We come back in more detail on the subject of small separations in Sect.~4, as they present a special behaviour in models 7 and 8.

\subsection{Seismic analysis of the accretion models 5 and 6}

Two models have been computed with the assumption that the observed overmetallicity was due to accretion of hydrogen poor matter onto the star during the early phases of planetary formation. Both models have a solar metallicity in their interiors while model 5 (AC) has [Fe/H] = 0.23 and model 6 (AC) [Fe/H] = 0.27 in their outer layers. The external and internal parameters of these models are given in Tables~\ref{tab2} and \ref{tab3}. Both models have a convective core which is smaller than the ones obtained for the overmetallic models with the same atmospheric metallicity (models 3 and 4). 

The large separations, echelle diagrams and small separations for these models are given in Fig.~\ref{fig2}, \ref{fig3} and \ref{fig4} (lower panels). Although these models show no qualitative differences with the overmetallic models, due to the fact that they all have convective cores, they do present visible quantitative differences, which are due to the different sizes of their convective cores. The situation here is not the same as for stars with smaller masses, for which the accretion models, which have no convective cores, follow evolutionary tracks very different from those of the overmetallic models (see Bazot \& Vauclair \cite{bazot04}). We expect however that the quantitative differences between these models should be detectable at the precision of the CoRoT observations.

\section{The special cases of models 7 and 8 : negative small separations}

Models 7 and 8 correspond to the Takeda et al. (\cite{takeda05}) error boxes in the log $g$-log $T_{eff}$ diagram (Fig.~\ref{fig1}, upper panels, left for model 7, right for model 8). Model 7 is at the end of the main-sequence, model 8 at the beginning of the subgiant branch. 

\subsection{Seismic analysis of models 7 and 8}

These models present very interesting features, clearly visible in Fig.~\ref{fig5} : the $\ell=0$ and $\ell=2$ lines cross at frequencies of 2.3 mHz for model 7 and 3.15 mHz for model 8. Meanwhile the $\ell=1$ and $\ell=3$ lines become very close at large frequencies. In the lower panel, we may check that the small separations between $\ell=0$ and $\ell=2$ become negative at the crossing frequencies.

These characteristic features are related to the rapid variations of the sound velocity near the helium-rich cores (Fig.~\ref{fig6}, upper panels). In model 7, the core is convective, with a high helium abundance Y = 0.876. In model 8, which corresponds to the beginning of the subgiant branch, the central convection has disappeared but a sharp edged helium core remains. 

While the $\ell=0$ modes propagate inside the whole star, from the surface down to the centre, the $\ell=2$ modes turn back at a layer whose radius (turning point) is given by :
\begin{equation}
S_l(r_t) = \omega = 2 \pi \nu
\end{equation}
where $\nu$ is the frequency of the considered mode and $S_l(r)$ the Lamb frequency, given by :
\begin{equation}
S_l(r) = \frac{c(r)}{r} \sqrt{l(l+1)}
\end{equation}

We have computed the turning points $r_t(\nu)$ for the $\ell=2$ modes in models 7 and 8, that is the radii for which:
\begin{equation}
c^2(r_t) = \frac{2 \pi^2 \nu^2}{3 } r_t^2
\end{equation}

The results are given in Fig.~\ref{fig6} (lower panels). For both models, the turning points of the $\ell=2$ waves reach the edge of the core at the exact frequency for which the $\ell=0$ and $\ell=2$ lines cross in the echelle diagrams. 

\subsection{Negative small separations : discussion}

For the present study, the p-mode frequencies have been computed in an exact way, using the code PULSE (Brassard et al. \cite{brassard92}). The fact that the computed small separations change sign for models with helium cores is an important result. Negative small separations are at first sight in contradiction with the so-called ``asymptotic theory'' developed by Tassoul (\cite{tassoul80}). We show below that this behaviour may be understood if we relax one of the Tassoul's assumptions, namely if we take into account the fact that the $\ell=2$ waves do not travel down to the centre of the star, below their turning point.
In the usual asymptotic theory, the small separations may be written :
\begin{equation}
\delta \nu = \nu_{n,l} - \nu_{n-1,l+2} \simeq -(4l+6)\frac{\Delta \nu }{4\pi ^{2}\nu _{n,l}}\int_{0}^{R}\frac{1}{r} \frac{dc}{dr}dr
\label{eq4}
\end{equation}
which is always positive. 

However, this expression is obtained with the assumption that the integrations may be performed between $r=0$ and $r=R$ for small values of $\ell$ as well as for $\ell$=0, while they should be computed between $r_t$ and $R$, where $r_t$ is the internal turning point of the waves. 

In case of helium rich cores, the sound velocity gradient which enters the integral in equation~(\ref{eq4}) strongly varies near the centre, as shown on Fig.~\ref{fig7}, upper panels. As a consequence, the integral $I(r)$, defined as :
\begin{equation}
I(r)=\int_{r_{t}}^{R}\frac{1}{r}\frac{dc}{dr}dr
\end{equation}
changes drastically, so that $I(r)$ is very different from $I(0)$ for small values of $r$ (Fig.~\ref{fig7}, lower panels).
 
To understand how the small separations may become negative, let us go back to the approximate computations of the frequencies proposed by Tassoul (\cite{tassoul80}), in which we reintroduce integrals between $r_t$ and $R$ instead of zero and $R$.
The approximate expression for $\nu _{n,l}$ may be written:
\begin{equation}
\nu_{n,l}\simeq (n+\frac{l}{2}+\frac{1}{4}+\alpha )\Delta \nu_l -\frac{l(l+1)\Delta \nu_l }{4\pi ^{2}\nu _{n,l}}\left[ \frac{c(R)}{R}-\int_{r_{l}}^{R}\frac{1}{r}\frac{dc}{dr}dr\right]- \delta \frac{\Delta \nu_l^2}{\nu _{n,l}}\ 
\end{equation}
which is the usual Tassoul's expression where the lower limit for the integral $r_l$ is the turning point for the $\ell$ mode with frequency $\nu _{n,l}$; $\Delta \nu_l$ is the large separation, generally approximated as $\Delta \nu_0=\frac{1}{2t_a}$ where $t_a$ is the acoustic radius of the star; $\alpha$ corresponds to a surface phase shift and $\delta$ is a function of the parameters of the equilibrium model. For the following discussion, we keep $\Delta \nu_l$ in the equations as a working hypothesis instead of $\Delta \nu_0$, which means that we allow $\Delta \nu_l$ to be different for different $\ell$ values.

Under these assumptions, neglecting the ratio $\frac{c(R)}{R}$ in front of the integral and the $\delta$-term, the small separations between $\ell=0$ and $\ell=2$ may be written :
\begin{equation}
\delta \nu_{02}\simeq (n+\frac{1}{4}+\alpha )(\Delta \nu_0-\Delta \nu_{2})-\frac{3 \Delta \nu_2 }{2\pi ^{2}\nu_{n,l}} I(r_t)   
\end{equation}

For model 8, $I(r_t)$ changes sign when $r_t$ reaches the edge of the helium core (Fig.~\ref{fig7}, lower panel, right). In this case, the reason why $\delta \nu_{02}$ changes sign is evident. For model 7 (left), the integral becomes close to zero without changing sign. However, a close look at the large separations (Fig.~\ref{fig5}, upper left) shows that, in the considered range of frequencies, $\Delta \nu_2$ is systematically larger than $\Delta \nu_0$ by about $0.2 -0.3$ $\mu$Hz, which represents a signature of the fact that the propagation times for the $\ell=2$ waves is smaller than that of the $\ell=0$ waves. This behaviour is logically related to the drop of the sound velocity in the helium core. With characteristic values of $n$ of order 30, a rapid computation shows that such a difference in the large separations is enough to induce a change of sign for $\delta \nu_{02}$ when the turning point of the waves reaches the edge of the helium core.

This is an important result which will be discussed in more details for the general case of solar-type stars in a forthcoming paper.

\section{Summary}

The exoplanet-host star HD~52265 is one of the main targets of the CoRoT satellite and, as such,  will be observed as a ``long run'' during five consecutive months. This solar-type star has one known planet, a ``hot Jupiter'' orbiting at 0.5 AU with a period of 119 days. As most exoplanet-host stars, its measured metallicity is larger than that of the Sun. From various observing groups, the determined values range between [Fe/H] = 0.19 to 0.27. Each of these metallicities is given with consistent spectroscopically-obtained values of gravity and effective temperature.

We performed an extensive analysis of this star in the same way as we did for the exoplanet-host star $\iota$ Hor (Laymand \& Vauclair \cite{laymand07}). We first presented a review of the spectroscopic observations: five groups of observers gave different sets of values for its metallicity, gravity and effective temperature. We then computed eight realistic models of HD~52265, each of them consistent with at least one of these sets of observed parameters. They present oscillation frequencies and seismic signatures (large and small separations, echelle diagrams), with differences much larger than the precision expected with CoRoT. The differences in the large separations among our models vary from 4 to 25 $\mu$Hz. For the small separations, they vary from 1 to 4 $\mu$Hz. Since the 150 days long runs of CoRoT will provide frequencies with an uncertainty of the order of 0.1 $\mu$Hz, it should be possible to determine unambiguously which of all the potential models gives the best fit of the observations.

The computed models which correspond to the Takeda et al. (\cite{takeda05}) atmospheric parameters, present a special behaviour. Two kinds of evolutionary tracks can cross the error box. For one of them, the star is at the end of the main sequence (model 7) and it has a helium-rich convective core. For the second case, the star is at the beginning of  the subgiant branch: the convective core has disappeared but a helium core remains, with a sharp boundary. For both models, the small separations become negative at a particular frequency related to the penetration of the acoustic waves inside the core. We have given a preliminary discussion of this important behaviour, which will be discussed in more detail for the general case of solar type stars in a forthcoming paper.

The detailed modelling of HD~52265 shows that seismic tests are more powerful than classical spectroscopy for determining the external stellar parameters.  It also leads to more precise values of the mass and age of the star, and gives constraints on its internal structure. 
With asteroseismology, we are entering a new era for the study of the structure and evolution of solar-type stars. Observing and studying the oscillations of exoplanet-host stars compared to those of stars without detected planets is extremely interesting in view of determining their differences in internal structure and chemical composition. This subject is closely related to the studies of planetary formation and evolution, and to the process of planet migration towards the central star.

\end{document}